# Development of a threat modelling framework and a web-based threat modelling tool for micro businesses

by

Etkin Getir

A Dissertation

submitted for the degree of Master of Science in

Computer Science

University of Essex Online

June 2024

# 1. Introduction:

In today's world where cyber security threats are more prominent than ever, threat modelling emerges as an essential undertaking for individuals with an online presence and who are concerned about their security and privacy (Braiterman et al., N.D.). In essence, threat modelling is a systematic approach of identifying potential threats to a valuable asset, such as a system, organisation, process, or software, that needs to be protected from cyber security risks.

Threat modelling usually encompasses developing a deeper understanding of the asset by decomposing it and drawing a diagram depicting its elements and the associations (i.e. data or control flow) between them, reviewing each element, association, or the system as a whole from cyber security perspective and identify the potential threats to the asset in question. The resulting threat model helps the stakeholders to devise mitigations against the threats that are identified.

Although its origins in information security context can be traced back to the 1970s (Cummings et al., 2023), threat modelling gained prominence in the cyber security and software development community in the late 1990s, primarily due to research sponsored by the NSA and DARPA which resulted in the attack trees (Salter et al., 1998), and also the development of a threat modelling framework by Microsoft cyber security professionals (Kohnfelder & Garg, 1999). Over time, many threat modelling frameworks have been introduced, each with a different approach and/or focus, and threat modelling still continues to be a major area of research. However, each threat modelling framework has one thing in common: They require a person or a team with a certain level of technical expertise and understanding to apply the framework and enumerate the threats.

Nevertheless, it is not always possible to find the required competence in all kinds of organisations that face cyber security risks. Even though large organisations should have no problem assembling a team of experts to lead the threat modelling process, at the other end of the company size spectrum, the micro businesses (MBs) are rarely able to do the same.

MBs, which are a subset of small and medium-sized enterprises (SMEs), are defined in a number of different, but similar ways. While the U.S. Small Business Administration informally defines a MB as having fewer than 10 employees (Office of Advocacy, 2017), the European Commission's (EC) definition is more precise: The EC defines MBs as enterprises employing fewer than 10 persons and having a turnover of less than EUR 2 million (European Commission, 2020). On the other hand, in the UK, companies with 10 or fewer employees and an annual turnover of less than £632,000 are categorised as "micro" business (Companies House, 2023).

The challenges MBs face in terms of threat modelling are twofold: First and foremost, MBs are understaffed, so it is usually not possible to designate an employee to oversee the cyber security operations of a business. In fact, most MBs are sole traders. The UK Office for National Statistics (Office for National Statistics, 2021) estimates that there were roughly 2.5 million MBs in the nation as of 2021, employing 4.9 million people. The average number of employees per company was therefore 1.96. This suggests that it is not possible to handle cyber security professionally in MBs. Secondly, because these businesses are financially constrained, it is not realistic to expect them to invest in cyber security. It is therefore unlikely that they would hire a cyber security professional to model the threats to their business or transfer the risk with the help of third parties, the latter also partly due to the lack of cyber security knowledge among MB stakeholders.

However, despite their small size, MBs account for a significant part of the economy and their impact is by no means small at all. In 2022, 95% of all businesses in the UK were MBs. These businesses were also responsible for 19% of total turnover and 32% of total employment in the country (Hutton, 2022). Similarly, in the EU, MBs made up 94.1% of all businesses in 2023, generating 16.6% of total turnover and employing 30.1% of the total workforce (Eurostat, 2023).

Yet, regardless of their significant footprint, there is little research that focuses on how cyber security, and specifically threat modelling, is handled in MBs. The aim of this study is to fill this gap by exploring the challenges of threat modelling for MBs. In addition, it seeks to propose a non-technical threat modelling framework and develop a web-based threat modelling tool based on this framework. To this end, the study will attempt to answer the following research questions:

RQ1. Are there any existing threat modelling frameworks applicable to the MBs?

RQ2. Are there any relevant aspects of the existing threat modelling frameworks that are relevant to the MBs?

RQ3. Is it possible to design a non-technical threat modelling framework so that MB owners can model the cyber security threats they face in the absence of cyber security expertise and knowledge?

The EC definition of MBs was adopted for the purposes of this study for two reasons: Firstly, it is more inclusive, so that the study would benefit from a larger sample size during the survey. Secondly, the sample was not drawn exclusively from the UK but rather, from around the world and predominantly from European countries.

The rest of the paper is structured as follows: In the next section, a literature review is conducted to identify threat modelling frameworks and their relevance to MBs. The research methodology and implementation are elaborated in the third and fourth sections respectively. The penultimate section evaluates the results of the research. The paper ends with conclusions.

## 2. Literature Review:

The literature review aims to identify the primary threat modelling frameworks, their relevance to MBs, and the current state of research about threat modelling in MBs. As such, the literature review is divided into three subsections: First, it provides an overview of the different threat modelling frameworks available in the literature, such as STRIDE, DREAD, and OCTAVE. This section examines the characteristics of each framework. The hybrid threat modelling methodologies, which rely on more than one of the primary threat modelling frameworks like Hybrid Threat Modeling Method (Mead et al., 2018) or Quantitative Threat Modeling Method (Potteiger et al., 2016), were left out of this literature review because the ultimate aim of the study is to identify relevant aspects of existing threat modelling frameworks for MBs, and the hybrid approaches do not provide any novel insights or unique perspectives for this specific context. Additionally, the exclusion of hybrid methodologies allows for a more focused analysis of each framework and its applicability to MBs, avoiding unnecessary duplication of findings caused by the core methodologies underlying these hybrid approaches, which are already analysed individually. The literature review then investigates the current state of research on threat modelling in MBs, highlighting gaps in knowledge. Furthermore, it explores the challenges and obstacles encountered by MBs when implementing threat modelling practices.

## 2.1. Threat modelling frameworks

Robert Barnard's attack profile, created in 1988, was the first attempt to represent threats in a computational context (Grigoriadis, 2019). However, it was not until the mid-1990s that threats were represented in a structured way.

Edward Amoroso used "attack trees" for the first time in 1994 to visually represent system threats (Amoroso, 1994). Five years later, Bruce Schneier further developed the concept of attack trees, providing a method for presenting threats in a visual and hierarchical manner (Scheiner, 1999; Schneier, 2015). The attack trees provided a way to analyse and prioritise potential threats based on their dependencies and relationships. This allowed for a more systematic approach to threat modelling, facilitating communication between different stakeholders involved in the security process and, perhaps more importantly, bridging the gap between technical details and strategic decision-making, making it easier for organisations to understand the potential impact of different threats.

The first formalised threat modelling framework, STRIDE, was developed by Loren Kohnfelder and Praerit Garg (1999) at Microsoft. STRIDE is a mnemonic, with each letter representing a threat category: Spoofing, Tampering, Repudiation, Information Disclosure, Denial of Service, and Elevation of Privilege. It provided a structured methodology for identifying and assessing threats, helping organisations to proactively identify vulnerabilities and develop effective countermeasures (Shostack, 2014). The main drawback of STRIDE is that it focuses primarily on technical threats and may not adequately address social engineering or human factors which have gained prominence in recent years. Furthermore, it necessitates a technical understanding and expertise to implement it effectively. Despite these limitations, STRIDE remains a valuable and popular approach in the field of threat modelling.

| Type | Name | Focus | Point of view | Technicality | Characteristics |
|---|---|---|---|---|---|
| Threat Modelling Framework | STRIDE | Identifying threats in software systems | Developer | High | Most established framework. Requires deep knowledge of the whole system. |
| | DREAD | Risk assessment of security threats | Security Analyst | Medium | Used to assess the severity of the risk. |
| | LINDDUN | Privacy threats | Privacy Specialist | Medium | Focuses on data protection and privacy issues. |
| | SDL | Integrating security practices into software development | Developer | High | Designed by Microsoft for software development teams. |
| | OCTAVE | Organizational risk management | Organization | Low to Medium | A framework for managing information security risks. Emphasises organisational risk over technical risk. |
| | OCTAVE-S | Smaller organizations | Small Business | Low | Adaptation of OCTAVE for small-scale businesses. |
| | TRIKE | Risk management tool | Security Analyst | Medium | Lacks documentation. Emphasis on automation. |
| | VAST | Large-scale enterprise systems | Enterprise | Medium to High | The framework that ThreatModeler (paid) tool is based on. Supports Agile DevOps. |
| | PASTA | Risk-based threat modelling | Risk Manager | High | Seven-stage process, focusing on attack simulation. |
| | Persona non grata | Understanding attackers' motivations and tactics | Behavioural Analyst | Medium | Personifies potential attackers. |
| | Misuse Cases | Identifying and analysing potential misuse or abuse scenarios | System Designer / Developer | Medium to High | Emphasizes on how a system can be misused, focusing on negative scenarios to enhance security measures. |
| Incident Analysis Framework | Cyber Killchain | Identifying stages of a cyber attack | Incident Responder | Medium | Outlines stages from reconnaissance to actions on objectives. |
| | Diamond Model | Analysing cyber intrusions and threats. | Incident Responder / Security Analyst | High | Focuses on the interaction between four core elements (adversary, capability, infrastructure, and victim) to understand the cyber threats and how to respond to them. |
| Tools | CVSS | Assessing the severity of software vulnerabilities | Security Analyst | Medium | A framework offering a standardised method for quantitively assessing the severity and impact of a vulnerability, facilitating data exchange between security teams. |
| | Security Cards | Facilitating broad and creative thinking about computer security threats. | Security Analyst / Educator | Medium | A brainstorming toolkit for computer security threats across four dimensions. |
| Knowledge Base | MITRE ATT&CK | Understanding and cataloguing the strategies and methods employed by adversaries. | Security Analyst | High | Provides a matrix of tactics and techniques used by threat actors in the enterprise environment. |
| | MITRE CAPEC | Cataloguing common attack patterns and techniques that adversaries use to exploit known vulnerabilities. | Security Analyst / Developer | High | Provides a comprehensive database of attack patterns along with detailed descriptions to help in the identification and mitigation of vulnerabilities. |
| | MITRE CVE | Documenting publicly disclosed cyber security vulnerabilities. | Security Analyst/IT Professional | High | Industry standard for referencing to publicly known vulnerabilities and exposures. |
| Attack Trees | Attack Trees | Visualising the potential routes an attacker may follow to accomplish a hostile objective against a system or valuable resource. | Security Analyst / Systems Engineer | Medium to High | Provides a hierarchical diagram of attack vectors, allowing for systematic analysis of threats and vulnerabilities. |

*Table 1: Comparison of threat modelling methodologies, knowledge bases and tools*

Later, Microsoft developed the DREAD framework (Howard & LeBlanc, 2004) to complement the STRIDE framework. DREAD stands for Damage Potential, Reproducibility, Exploitability, Affected Users, and Discoverability. This framework allows organisations to prioritise and quantify the severity of identified threats based on these five factors. However, DREAD is criticised for its subjective nature and lack of specific guidance on how to assess each factor, and it is no longer recommended or used by Microsoft (LeBlanc, 2007; Shostack, 2014).

STRIDE was designed with software development in mind and was not intended to be used throughout the organisation. To address this shortcoming, OCTAVE (Operationally Critical Threat, Asset, and Vulnerability Evaluation) was developed as a comprehensive risk assessment methodology that can be applied across an entire organisation (Alberts et al., 2003). In contrast to technology-focused methodologies, which tend to concentrate on tactical issues, OCTAVE distinguishes itself by analysing organisational risk and concentrating on strategic, practice-related matters. The three-phase OCTAVE process starts with the analysis team determining the critical assets of the organisation as well as their security requirements and potential threats to them. This is followed up by the evaluation of information infrastructure. The third phase involves the identification of threats to the critical assets and the development of a risk mitigation plan. A variant of the methodology, known as OCTAVE-S, is specifically designed to assess risks within the context of SMEs (Alberts et al., 2005).

Trike is a security auditing framework developed in the mid-2000s (Saitta et al., 2005). Its main promise is the automation of repetitive tasks and its defensive perspective as opposed to the traditional offensive approach of other frameworks. In Trike, threat modelling starts with creating the requirements model, in which assets, actors, rules and intended actions are defined, and an Actor-Asset-Action Matrix is created based on CRUD actions. Then, the threats are generated and analysed. However, despite its initial promise, Trike did not gain widespread adoption, and its lack of documentation and relative complexity proved to be significant obstacles to its implementation.

Privacy stands out as a feature that is often overlooked in threat modelling methodologies, as evidenced by STRIDE's emphasis on "non-repudiation". In this context, LINDDUN was the first framework to address privacy threats. LINDDUN is also a mnemonic that stands for Linkability, Identifiability, Non-repudiation, Detectability, Disclosure of Information, Content Unawareness, Policy and Consent Noncompliance (Deng, 2010). One standout feature of the methodology is its emphasis on "repudiation" as a privacy feature as opposed to "non-repudiation" as a security feature in STRIDE. It also takes the concept of hard and soft privacy further by defining hard and soft privacy properties. With its focus on privacy, LINDDUN fills an important gap in existing threat modelling methodologies and can enhance an organisation's overall security posture. However, because it focuses solely on privacy, it can be used as a supplement to other threat modelling methodologies that address security concerns.

VAST (Visual, Agile, and Simple Threat Modelling) is a methodology developed in response to the rise of agile methodologies in the software development industry. It supports Agile DevOps and is built on three pillars: automation, integration, and collaboration (ThreatModeler, 2018). Furthermore, scalability is one of the framework's standout features. VAST is based on a commercial threat modelling tool called ThreatModeler and incorporates visualisation techniques to identify and analyse potential threats, making it a valuable tool for organisations looking to enhance their overall security posture. However, VAST's reliance on the ThreatModeler tool may pose challenges for organisations that do not have access to or familiarity with this specific tool. Additionally, the complexity of implementing VAST may require additional training and resources for teams to effectively utilise its capabilities.

PASTA (Process for Attack Simulation and Threat Analysis) is a risk-centric, 7-step threat modelling methodology (Ucedavélez & Morana, 2015). The process involves key stakeholders from outside the security and compliance teams and results in the enumeration of risks, mitigations, and a risk management strategy. Unlike VAST, PASTA does not rely on a specific commercial tool, making it more accessible to a wider range of organisations. Nevertheless, due to the laborious nature of the process, the successful implementation and integration of PASTA into current security practices may necessitate the allocation of dedicated resources and training.

Threat modelling is an activity that benefits all types of organisations. However, the technicality typically associated with threat modelling practices makes it overwhelming for people without a background in cyber security. In an attempt to simplify the threat modelling process and make it more accessible to a wider audience, the Threat Modelling Manifesto emphasises the need for everyone to engage in threat modelling and reduces threat modelling to four simple steps: "(1) What are we working on? (2) What can go wrong? (3) What are we going to do about it? (4) Did we do a good enough job?" (Broilerman et al., N.D.).

In addition to threat modelling frameworks that help to identify and prioritise potential security risks, there are also additional tools and sources at threat modellers' disposal. For instance, MITRE ATT&CK (Adversarial Tactics, Techniques, and Common Knowledge) is a library that categorises adversary tactics and techniques and provides detailed descriptions for each, including suggested detection methods and mitigations (Xiong et al., 2022; Georgiadou at al., 2021). MITRE also maintains CAPEC (Common Attack Pattern Enumeration and Classification), which provides an application security-focused catalogue of common attack patterns and their associated characteristics (Shaked, 2024). CVE (Common Vulnerabilities and Exposures), also maintained by MITRE, is another widely used resource in the cyber security community. CVE is a knowledge base of publicly known information security vulnerabilities and exposures, providing detailed descriptions and references to patches or mitigations (Branescu et al., 2024). By referencing CVE entries, organisations can remain abreast on the most recent vulnerabilities and implement the necessary measures to protect their systems and networks. In a similar manner, in the National Vulnerability Database (NVD), NIST brings together all the vulnerabilities in the CVE list with additional analysis (NIST, 2023). OWASP Top 10 is another important source that draws the attention of the cyber security professionals to the most severe risks for web applications (Mangaoang et al., 2024).

On the other hand, CVSS (Common Vulnerability Scoring System) is a framework that provides a standardised method for assessing the severity of vulnerabilities (FIRST, 2023). Currently at version 4.0, CVSS provides an intuitive interface for the calculation of a vulnerability score based on a number of metrics (i.e., base, supplemental, environmental, and threat metrics). Even though it aims to standardise the calculation of vulnerability severity, CVSS has its limitations. One limitation is that it relies on subjective input from users to determine the values of certain metrics, which can lead to inconsistencies in scoring. Furthermore, The CVSS does not consider the potential impact of a vulnerability on specific industries or organisations, which limits its suitability for individual needs (Figueroa-Lorenzo et al., 2020). Additionally, it is criticised by many for being too vague in calculation (Howland,

2023; Siddique, 2021) .

Two tools that could be useful in the threat modelling process, especially for those without technical knowledge, are Persona non Grata and Security Cards. Persona non Grata allows for the creation of fictional personas to simulate potential attackers (Cleland-Huang, 2014; Mead et al., 2017), while Security Cards is a "brainstorming toolkit" that drives those involved in threat modelling to think about potential threats (Denning et al., 2013). These tools, when used in conjunction with other threat modelling methodologies, help to ensure comprehensive coverage of potential risks, as in the case of Hybrid Threat Modelling Method (Mead et al., 2018).

## 2.2. Cyber security challenges faced by MBs

Even though they seem to have a relatively smaller attack surface, MBs find themselves frequently victims of different types of cyber-attacks. Indeed, in 2023, 10% of MBs experienced a cyber-crime and 31% of the MBs have identified attacks or breaches (UK DSIT, 2023) and in 2024 the latter figure climbed to 47% (UK DSIT, 2024). Considering the poor detection mechanisms in the MBs, it can be argued that the real number of attacks and breaches is likely much higher. Moreover, as organisations get smaller, the average cost of a data breach per employee tends to be higher than in larger organisations (IBM Security, 2023). In a similar fashion, the adoption rate of digital security measures in smaller organisations is relatively low compared to larger ones (OECD, 2021).

Budgetary constraints stand out as a major obstacle for MBs in implementing robust cyber security measures (Kabanda et al., 2018). Initial and maintenance costs of cyber security controls are often viewed as too expensive for SMEs, leading to their reluctance to adopt such measures (Pawar & Palivela, 2022) and this is especially true for MBs due to their smaller budgets. Indeed, MBs, especially the single-person companies are among those with the lowest annual cyber security spending (Wilson et al., 2023).

As the smallest organisations in the business spectrum, MBs are often understaffed, which often leads them to adopt a DIY approach to security (Berry & Berry, 2018), attempting to address cyber security needs through the business owner or an existing member of the business (Osborn & Simpson, 2018) who are not capable of devising a cyber security strategy or applying the necessary security controls.

Another important obstacle for MBs is their general low level or lack of cyber security knowledge (Cartwright et al., 2023; Alahmari & Duncan, 2020). Some MB owners even find simple solutions like using password managers to create and store strong and different passwords for different services difficult to implement (Jamil, 2022). The lack of knowledge has a number of consequences for MBs: First and foremost, not knowing the threat landscape and the cyber security tools at their disposal makes them vulnerable to any threat that may target them. In addition, if they decide to pursue a cyber security strategy, they do not know where to start (UK DSIT, 2023) and if they opt to ask for external help, they do not know what to ask for (Cartwright et al., 2023). Similarly, this knowledge gap prevents them from benefiting from knowledge bases, threat intelligence platforms, or government sources to keep informed about the latest threats and mitigations because they either do not know what to look for or the content is too technical for them.

Lack of knowledge inevitably leads to a lack of awareness of cyber security. When they are unaware of the significance of cyber security and their own level of readiness, MB owners tend to adopt an optimistic view of their own cyber security posture (Wilson et al., 2023) and fall back on the misconception that the company's assets are not valuable enough to be a lucrative target for attackers, especially when compared to those of large organisations (Hoppe et al., 2021; Sangani & Vijayakumar, 2012). In this case, when they allocate the scarce resources they have, they prioritise the tasks that are directly related to the business goals, leaving cyber security as an afterthought (UK DSIT, 2023; Osborn & Simpson, 2018).

An organisation's cyber security cannot be thought of without considering the behaviour of the people involved. As a matter of fact, an organisation's overall cyber security level is heavily influenced by the human factor, and inappropriate employee behaviour is a significant cause of cyber security breaches (Alahmari & Duncan, 2020; Hoppe et al., 2021; Wilson et al., 2023; Caldwell, 2015). In addition, the time constraints inherent in MBs, where most employees are multi-tasked or most tasks are taken care of by MB owner or their family members (Berry & Berry, 2018), make the display of unsafe behaviour more commonplace (Chowdhury et al., 2020), and their behaviour tends to worsen under 'security-related stress' caused by the perceived complexity of security mechanisms (D'arcy et al., 2014). It should also be noted that many MB owners exclude employees from the organisation's cyber security strategy, exacerbating the problem (Patterson, 2017) and hindering the building of a security culture (Hoppe et al., 2021).

Without the required expertise, MBs have difficulty when they consider investing in cyber security in the absence of a roadmap and clear guidelines (Pawar & Palivela, 2022). Nevertheless, in recent years, there have been many government-led initiatives to remedy the situation and provide small organisations with straightforward guidance and suggestions, such as NCSC's Small Business Guide (NCSC, 2018) or NIST's Small Business Cybersecurity Corner (NIST, n.d.). However, these guides may still require a basic level of knowledge and may fail to address the unique characteristics of MBs, which are typically comprised of a small group of people, if not only the MB owner. And on some occasions, MB owners express their doubts about sources of information or cyber security support they received from government agencies or other stakeholders (Tam et al., 2021; Yudhiyati et al., 2021).

The evolving technology landscape requires new competencies. For example, migration to the cloud is gaining momentum among MBs, but it also brings new challenges (Nagahawatta et al., 2021). MBs now need to understand what risks are being transferred to the cloud service provider, what capabilities are available to them in the cloud, and how to integrate the cloud capabilities with their traditional aspects of the business. However, most MBs still lack the technical expertise to use cloud services effectively and improve their cyber security posture (Tam et al., 2021).

Furthermore, the complexity and diversity of industry standards (Tam et al., 2021) and the financial implications of transitioning from legacy systems (Osborn & Simpson, 2018) represent additional challenges that MBs must navigate.

**2.3. Threat modelling methodologies, tools, and sources available to the MBs**

Despite the plethora of threat modelling and risk management frameworks, each of them necessitates a certain level of technical knowledge and/or collaboration between multiple stakeholders with different expertise. However, as outlined above, MBs are characterised by the low number of employees, the absence of a dedicated IT department and technical knowledge, and the prominent role that the business owner plays. For this reason, it is usually not feasible to utilise these methodologies in the context of MBs. The only risk management framework that is tailored specifically for small businesses is OCTAVE-S (Alberts et al., 2005), however, this methodology assumes that the company has between 20 and 80 employees and is still quite complex for non-technical people. Consequently, it is not a particularly relevant approach for MBs. Similarly, threat modelling tools such as OWASP Threat Dragon (Granata & Rak, 2024) or Microsoft Threat Modelling Tool (Microsoft, 2022) require knowledge of some of the threat modelling frameworks and some basic techniques such as creating data flow diagrams, and are therefore not suitable for MBs.

In response to this gap, some methodologies have been introduced by numerous researchers. One example is the Least Cybersecurity Control Implementation (LCCI), which is an approach that focuses on implementing the minimum number of cyber security measures (Pawar & Palivela, 2022). However, even though it helps SMEs to improve their cyber security posture, LCCI is fairly complex and therefore not suitable for MBs. However, it has the potential to become usable for MBs if it is accompanied by a tool with an easy-to-use, non-technical interface. Another similar work is the threat-based cyber security risk assessment methodology developed by van Haastrecht et al. (2021). Unlike LCCI, the methodology comes with an application with an intuitive interface that provides risk scores and recommendations. The risk indicator is a useful tool for SMEs, but a non-technical theoretical framework can help SMEs build a security culture and raise cyber security awareness. Finally, the Cybersecurity Evaluation Tool developed by Benz & Chatterjee (2020) is based on NIST's Cyber Security Framework, uses only 35 of the total 96 standards that are relevant to SMEs, and provides a scorecard with recommendations as a result of an online questionnaire.

Outside of academia, government agencies also provide guidance to SMEs on cyber security best practices. For example, in the UK, the National Cyber Security Centre (NCSC) offers a range of resources and advice specifically tailored for SMEs.

Among others, these resources include Cyber Essentials, a government initiative that aims to assist small and medium-sized enterprises (SMEs) in safeguarding against prevalent cyber threats (UK Government, 2023); Cyber Essentials Readiness Tool, which uses the answers of the users to an online questionnaire to provide customised suggestions with additional explanations (NCSC, 2023); and the Cyber Aware campaign to raise cyber security awareness among micro and small businesses (Rama & Keevy, 2023). These resources offer practical guidance on implementing basic security measures, but their effectiveness in influencing MB behaviour is doubtful (Cartwright et al., 2023). On the other hand, in the US, there are similar initiatives, such as the National Institute of Standards and Technology (NIST) Cybersecurity Framework, which offers organisations a collection of principles and recommendations to help them manage and reduce cyber security risks (NIST, 2024). The framework also has a version that is tailored for small manufacturers. Additionally, the Small Business Administration (SBA) and NIST offer resources and training programmes specifically tailored to help small businesses enhance their cyber security posture through the Small Business Cybersecurity Corner (NIST, N.D.). In the EU, ENISA leads similar projects, such as the Cybersecurity Guide for SMEs (ENISA, 2021) and SecureSME (ENISA, N.D.). Another notable project sponsored by the EU is SMESEC, which is a promising framework and web-based tool that is intended to help SMEs evaluate and enhance their cyber security measures (Cucurull et al., 2020; SMESEC Consortium, N.D.). Finally, ISO/IEC 27002 sets out standards for information security management systems at the international level (ISO, 2022). Designed to work in tandem with ISO/IEC 27001, it offers a full suite of controls and recommendations that businesses of any size can use. However, when it comes to implementing the recommendations in the standard, MBs may not be able to do it on their own due to their lack of basic cyber security knowledge and they may have to turn to compliance platforms and firms that impose costs beyond their budgets, and therefore they may not be tempted to do so.

## 3. Methodology

A qualitative, seven-step methodology was used to achieve the research objectives.

### 3.1. Literature review on the current state of research on cyber security in MBs

University of Essex Library, Google Scholar, and IEEE Library were the primary resources consulted for the literature review. The review focused on three key areas: Firstly, the existing threat modelling frameworks were reviewed, especially with regard to their focus and level of technicality. Then, the challenges faced by MBs in implementing cyber security measures were identified and analysed. Finally, the effectiveness of the cyber security tools, methodologies, and resources available to the MBs was assessed. This comprehensive literature review served as a strong basis for the creation of the first questionnaire, as well as for the subsequent stages that will involve the development of a threat modelling framework and a threat modelling tool tailored for MBs.

### 3.2. An initial qualitative survey to explore how cyber security is handled in MBs and the pitfalls they encounter

The second step involved conducting a qualitative survey to gain insights into the current practices and challenges encountered by MBs in the implementation of cyber security measures. This survey provided valuable information that informed the development of the threat modelling framework and tool specifically designed for the unique needs of MBs.

The questionnaire was administered online using Google Forms and targeted MB owners and managers. It included questions about their educational background and experience with cyber security, as well as details about their cyber security practices and the challenges they encounter. The initial questionnaire is presented in Appendix A.

A total of 320 MB owners and managers were contacted using their publicly available email addresses and/or LinkedIn profiles, out of which 21 responded to the questionnaire. The responses from this survey helped identify common patterns and issues that should be taken into account when developing the threat modelling framework and tool.

**3.3. Analysis of the initial survey results and identification of the problem areas**

In this phase, the initial survey results were analysed to identify the threat modelling experiences of MB owners and the problems they face in this area. The findings, which are discussed in Section 5, were used to design the framework and web-based tool in the next steps.

**3.4. Identification of threat modelling frameworks that are relevant to the MBs**

The threat modelling frameworks, tools and knowledge bases identified in the literature review were examined to determine whether there are aspects of these frameworks that can be used by MBs that do not have the necessary cyber security expertise. A summary of the findings is presented in the table below.

| Threat Modelling Framework / Tool / Knowledge Base | Relevant aspect |
|---|---|
| STRIDE | Relevant threat categories for MBs. |
| DREAD | The idea of prioritising threats based on their damage potential, reproducibility, number of users affected and ease of detection. |
| LINDDUN | The emphasis on privacy. |
| OCTAVE-S | Focus on organisational risk, security practices and self-management. Creation of asset-based threat profiles and critical review of infrastructure vulnerabilities. |
| CVSS | The idea of ratings and a rating system to represent the severity of a threat. |
| Persona non Grata and Security Cards | The abstraction of potential threat actors and threat patterns in a manner that is easily understandable, which can help to raise the MB owner's awareness of potential threats that they may not have been aware of previously. |
| Misuse Cases | Misuse cases can help the MB owner see how an intended use case can lead to a potential threat. |
| Data Flow Diagram (DFD) | DFDs can help MB owners to have a holistic view of their business. |
| Attack Trees | The idea of presenting threats in a visual and hierarchical way. |
| MITRE ATT&CK, CAPEC and CVE | An industry-aligned knowledge base can help MB owners understand the |

| | potential threats and mitigations. However, as these knowledge bases are highly technical, they require external help to interpret the technical aspects. |
|---|---|
| NSCS Cyber Essentials and NIST Small Business Cybersecurity Corner | Suggestions and action plans for small businesses. |

*Table 2: The relevant aspects of existing threat modelling frameworks, tools and sources to MBs*

Although the existing threat modelling frameworks are too technical for the average MB owner, as shown in Table 2, it is possible to identify some aspects of these frameworks that are either less technical or that would help an MB owner to understand the cyber threats to their business.

While the pioneering software development centric threat modelling framework STRIDE is not suitable for implementation in an MB environment, it is possible to relate its threat categories to those of MBs. Spoofing, Tampering, Repudiation, Information Disclosure, Denial of Service and Elevation of Privilege are all relevant threats to MBs but require non-technical interpretation and possibly rewording to make them more accessible to non-technical people. On the other hand, LINDDUN draws attention to the issue of privacy, which has recently gained prominence. Indeed, privacy considerations are particularly relevant for MBs that interact closely with their customers. In addition, most MBs have a highly variable and large customer base, sometimes with very low retention rates. For example, consider a restaurant that serves Chinese food: There will certainly be some loyal customers who prefer Chinese food and prefer the restaurant for a number of reasons, such as the quality of service or the speciality of the restaurant, but the majority of customers will be those who like to dine out at different restaurants, try new flavours and dine out only occasionally. If that restaurant wants to introduce an online or tableside ordering system that is facing the customer, or if the business decides to conduct a survey to understand customer preferences, the privacy concerns will be much greater than those of businesses with an invariable, permanent customer base.

The approach of the OCTAVE-S methodology is particularly valuable for MBs. First and foremost, it is based on a strategic, practice-based assessment, as opposed to the technology-focused approach of many other frameworks. It can be argued that an organisation's assets are the most reliable starting point for an MB's threat modelling practices, as they are more tangible and require less technical knowledge to assess. Phase 1 of the OCTAVE-S framework, 'Build Asset-Based Threat Profiles', aligns well with this argument. MBs also operate an IT infrastructure that encompasses their organisational assets, including the network (wireless or otherwise), facilities, data storage and so on. In this sense, the next stage of the framework, 'Identify Infrastructure Vulnerabilities', is also relevant to MBs, as it aims to identify the weaknesses in the organisation's infrastructure. Therefore, although OCTAVE-S is not directly applicable to MBs, as it requires a small team to implement and it is not possible for the MB owner to implement it alone, its approach is very insightful in terms of threat modelling at the MB level.

Once the threats have been identified, visualising them would help communicate an organisation's threat landscape to non-technical people and provide a holistic view of the organisation and the threats it faces. In this respect, the Data Flow Diagram (DFD) is a convenient way of depicting an organisation's assets, actors and, perhaps more importantly, the flow of data between them, all of which might otherwise be too abstract for an MB owner to understand. Similarly, attack trees can be used to represent threats in a hierarchical manner, allowing one to see which threats, individually or in combination, could lead to an undesirable situation.

Focusing solely on enumerating threats without prioritising them can be daunting for an MB owner, as they may end up with a long list of threats that could affect their business. However, on closer inspection, some threats may be of little significance to the business or have a low probability of occurring. This is where a risk assessment framework that measures the likelihood and impact of a threat, or a scoring system that better communicates the severity of a risk to non-technical people, can be helpful. Having carefully assessed the risks, an MB owner can decide whether to mitigate, transfer, accept or ignore the risk.

Without sufficient risk assessment experience, MB owners may find it difficult to identify possible threat actors or how their systems and assets might be misused. In particular, they may find it difficult to understand that a benign actor, such as a dedicated employee, may in some circumstances pose a threat to the organisation. In line with the persona non grata approach, this difficulty can be overcome by introducing MB owners to the archetypes of employees, customers, suppliers and other people with whom their business interacts, and providing more information about how each of these can put the business at risk. In addition, misuse cases, which can be easily derived from the intended use cases of the system, can give MB owners some indication of how their systems can be exploited and raise their awareness.

The benefits of knowledge bases are twofold: First, they provide threat intelligence and up-to-date information on the latest threats identified. They also go beyond this to suggest mitigations and safeguards against these threats. Although the technical nature of these knowledge bases makes it impossible for MB owners to make direct use of them, the usefulness of the information gathered here is undeniable. The main challenge is to translate the content of these knowledge bases into a language that MB owners can understand, and to integrate their content as much as possible into the tools to be developed for MBs.

In summary, in relation to RQ2, several aspects of existing threat modelling and risk management frameworks, knowledge bases and tools have been identified as relevant to threat modelling practices in MBs, as explained above and summarised in Table 2.

**3.5. Design a non-technical threat modelling framework based on the research done**

The objective of this phase was to develop a non-technical threat modelling framework for MB owners based on the findings of the literature review and the initial questionnaire. The threat modelling framework was intended to:

- Create a behavioural change by increasing MB owners' awareness of cyber threats and to get MB owners into the habit of applying mitigation measures by constantly modelling the threats to their business. By presenting a simple and understandable framework, MB owners would be able to keep it in mind in their day-to-day activities as well as when making important decisions about their IT assets or infrastructure.

- Serve as the foundation for a web-based threat modelling tool. The web-based threat modelling tool will put the threat modelling framework into practice and is intended to be used by MB owners to conveniently model threats by answering a few basic questions about their business. By using the tool, MB owners will gain a holistic view of their system and be able to enumerate the threats to their business, along with mitigation recommendations, with minimal effort and cost.

With these objectives in mind, a non-technical, asset-centric threat modelling framework was designed and formulated as a mnemonic so that it is easy to remember, adopt and use. The design details, considerations and the framework itself are presented in the Design and Implementation section.

**3.6. Design a web-based threat modelling tool based on the framework**

Based on the framework developed, a web-based threat modelling framework was developed. The primary objective of this phase was to provide the MB owners with a tool that would help them to implement the framework that had been designed in the previous step. The tool was designed as a web-based application rather than a desktop application to ensure cross-platform compatibility. The application outputs include a risk score, recommendations and a data flow diagram (DFD).

Detailed information on the design choices and development process is discussed in the Design and Implementation section and the test and validation results are presented in the Testing and Evaluation section.

**3.7. Follow-up survey on how MBs perceive the new tool and assess its impact on business cyber security**

Once the design of the threat modelling framework and the development of the web-based threat modelling tools had been completed, a follow-up survey was conducted using a questionnaire that included an introduction to the framework and the web-based tool, as well as several questions about the participants' experiences with both artefacts. The follow-up questionnaire is presented in Appendix B and the questionnaire results are discussed in the Testing and Evaluation section.

**4. Design and Implementation**

This section presents two resulting artefacts of the project. Firstly, the new threat modelling framework designed specifically for MBs is introduced, which also serves as the basis for the threat modelling tool. Then, the new web-based threat modelling tool is presented.

### 4.1. Threat Modelling Framework for MBs

#### 4.1.1 Features of the New Threat Modelling Framework

As a result of the literature review and initial survey, the need for a threat modelling framework specifically tailored to MBs was identified. Due to the technical nature of the available frameworks and the inherent challenges of MBs, it is evident that these frameworks are not suitable for MBs. Therefore, a simpler and more user-friendly approach is necessary.

In light of these findings, a new threat modelling framework was developed to address the unique needs and limitations of MBs. This framework aims to provide a practical theoretical basis for MB owners to effectively assess and mitigate potential threats to their operations, and to form the basis of a web-based threat modelling tool that will allow MB owners to easily input their information and generate a customised threat model. The key features of this new framework are as follows:

- **Non-technical**: As evidenced by the results of the initial questionnaire, the majority of MB owners do not have the necessary cyber security skills to engage in threat modelling. In this respect, a framework that would benefit MBs should not rely on the prior experience or technical expertise of MB owners, should be free of complicated technical explanations, and should provide concise, actionable recommendations.

- **Asset-centric**: While an asset-centric approach may not be ideal for the identification of threats under ideal conditions, as Shostack (2014) points out, asset-centric modelling helps non-technical people to contribute to the threat modelling process. In our case, MB owners typically do not have the technical expertise to use technology/software-centric or attacker-centric models, whereas they can identify the assets they want to protect. Therefore, assets provide a solid starting point for threat modelling in MBs.

- **Inside-out approach**: An inside-out approach to threat modelling in MBs involves first identifying threats and vulnerabilities pertaining to the MB owner themselves, and then working outwards to assess potential threats. The main reason behind this approach is the assumption that the most dangerous threat is the closest threat. For example, the MB owner's negligence regarding password policy has a greater impact on the organisation's cyber security than an employee's sloppy handling of their own password. By adopting an inside-out approach, MB owners can prioritise their security efforts and make informed decisions on improving the cyber security posture of their business, resulting in a more efficient and effective threat mitigation strategy tailored to the specific needs of the organisation.

- **Mnemonic**: In addition to providing the theoretical foundation upon which the web-based threat modelling tool is built, the primary goal of the framework is to create a behavioural change and habit of critically evaluating all components of their business. Therefore, it is imperative to capture the essence of the framework in an easily memorable mnemonic so that MBs can keep it in mind in their daily activities.

### 4.1.2 Introducing SEANCE Micro Business Threat Modelling Framework

SEANCE is a novel threat modelling framework tailored for MBs. It consists of six hierarchical layers, corresponding to the defence-in-depth approach. As the main purpose of the framework is to raise awareness and create a cyber security-oriented mindset among MB owners, it doesn't involve any calculations or filling in complicated spreadsheets. Instead, for each layer, MB owners are presented with key themes to consider at that layer, along with a non-exhaustive list of guiding questions designed to stimulate critical thinking and discussion about potential threats. Beyond the immediate benefits of adopting the framework, a side benefit would be to encourage them to delve deeper into cyber security concepts in the long term, so that they can improve their understanding of cyber security threats and mitigations, and further enhance their organisation's cyber security posture.

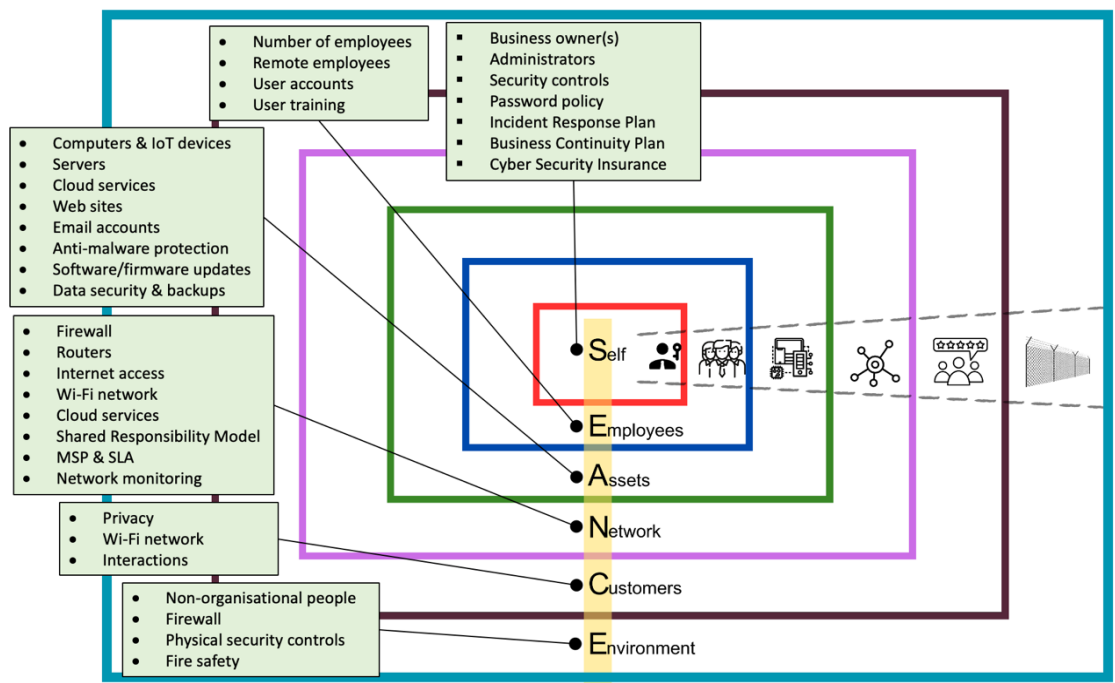

*Figure 1: Seance Framework and key themes at each layer*

The six layers of the SEANCE Framework are: (your)Self, (your) Employees, (your) Assets, (your) Network, (your) Customers, and (your) Environment. The inside-out approach is evident in the layered structure, which begins with the MB owner at the centre and ends at the external boundaries of the organisation. While some key issues are confined to a single layer, others appear at multiple layers. For example, firewalls appear at the network layer as a means of segregating the organisation's business network; they also appear at the environment layer as a means of filtering external traffic. The guiding questions at each layer are compiled and adapted from industry-standard sources such as the NIST Cybersecurity Framework (NIST, 2024) and the Cyber Essentials Readiness Tool (IASME Consortium, N.D.), among others.

In order to ensure that the framework is easy for MB owners to understand, a factsheet has been prepared (see Appendix D). The fact sheet explains the framework in a simple and non-technical manner, using a castle analogy. It also provides detailed information about the layers, including key themes and guiding questions, and includes external links to reputable sources for additional information and detailed explanation of some terms that may be difficult to understand.

Appendix C provides a comprehensive account of the SEANCE Framework, including detailed explanations, key themes, and guiding questions for each layer.

**4.2. Web-based Threat Modelling Tool Based on SEANCE Framework**

The web-based threat modelling tool is designed to complement the SEANCE framework by facilitating its implementation through a user-friendly interface and providing tangible outputs such as a DFD and actionable recommendations.

The tool is based on the Django Web Framework and relies on interaction with the user, who answers questions posed in six sections corresponding to the layers of the framework. Once all questions have been answered, the user is presented with an overall evaluation of their system, including a DFD and recommended actions to enhance their organisation's cyber defences.

The tech stack on which the web-based threat modelling tool is based is as follows:

- **Web Framework**: Django version 5.02
- **Programming Languages**: Python version 3.11, HTML5 and Javascript
- **Database**: SQLite version 3.45 (Development), PostgreSQL – Official Docker Image (Production)

- **Styling**: Bootstrap version 5.3, Custom CSS

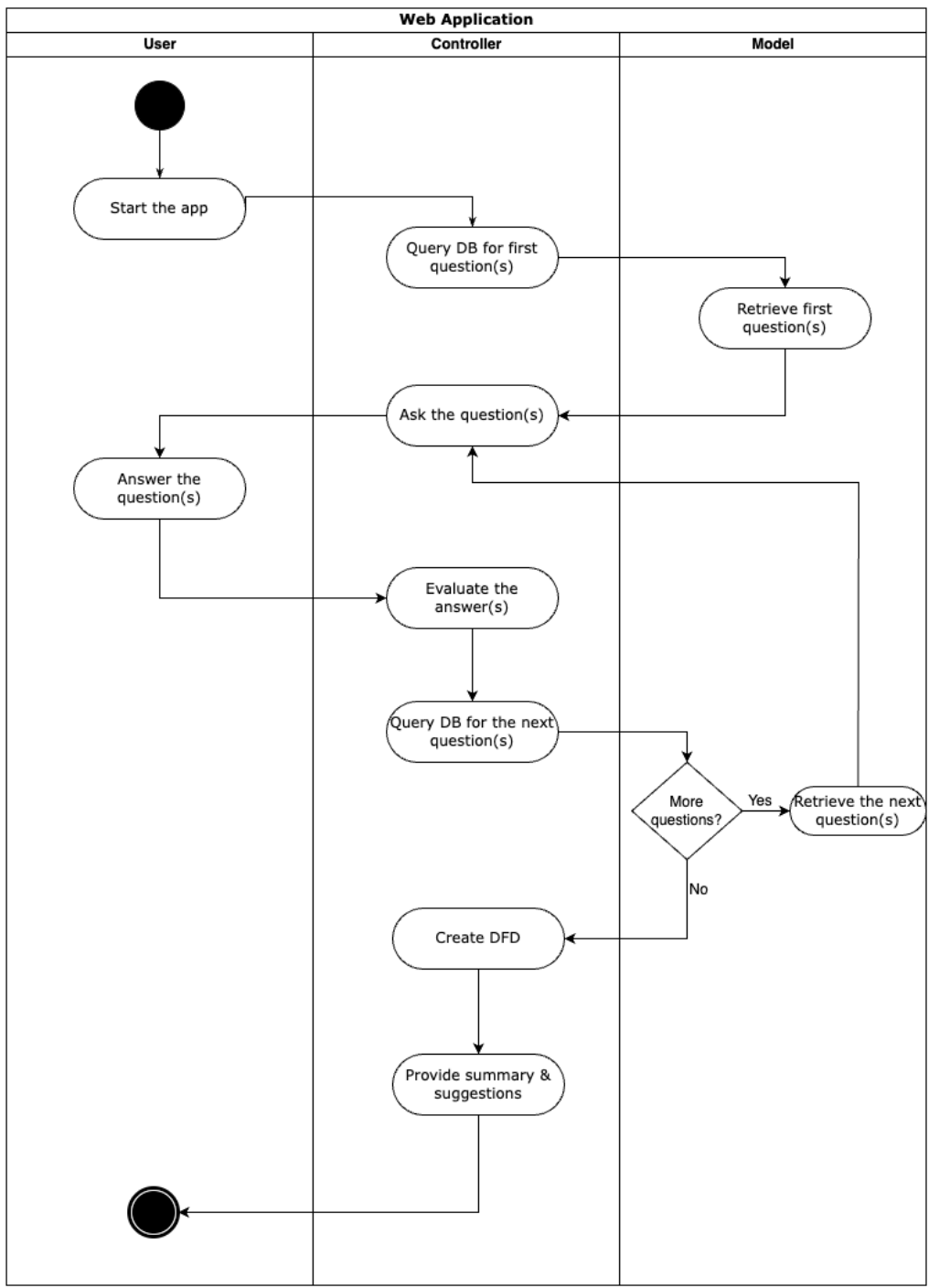

*Figure 2: Activity diagram of the web application*

### 4.2.1. Data handling and models

When designing the application's data handling architecture, a distinction was made between persistent and non-persistent data. The persistent data was stored in a database in multiple tables. During development, the database of choice was SQLite as it is a lightweight database supported by Django by default and an optimal choice for small applications. However, due to its performance limitations, the persistent data was migrated to a PostgreSQL database in production. On the other hand, non-persistent data was stored in data objects (i.e. instantiations of data classes) and local variables. The reason for this design choice was to improve the efficiency of the application by minimising CRUD operations on the database, as the Python data structures provide a more flexible way of storing non-complex data that is discarded when no longer needed.

The persistent data used in the application includes questions and possible suggestions. In the development phase, this data was stored in an SQLite database in separate tables. All questions, including main and sub-questions, are stored in a table with unique question IDs. Each question was also assigned a step, which was used by the application logic to retrieve the relevant questions depending on the user journey. Each record also has a Boolean field to indicate if it is a 'parent' (i.e. has children), the step where its children are, and a value that would help calculate the risk score depending on the user's answers.

| qid | question | qtype | section | value | children | parent | step | factor | dfd | pid |
|---|---|---|---|---|---|---|---|---|---|---|
| 10100 | Are you the sole owner of the bu… | 3 | 1 | 0 | 1.1.1 | 1 | 1.1 | 7 | 0 | 0 |
| 10101 | How many shareholders/partners d… | 4 | 1 | 0 | | 0 | 1.1.1 | 7 | 0 | 10100 |
| 10200 | Do you have and enforce a passwo… | 2 | 1 | 0 | 1.2.1 | 1 | 1.2 | 7 | 0 | 0 |
| 10201 | Does your password policy advise… | 1 | 1 | 0 | | 0 | 1.2.1 | 42 | 0 | 10200 |
| 10202 | Does password policy provide gu… | 1 | 1 | 0 | | 0 | 1.2.1 | 42 | 0 | 10200 |
| 10203 | Does password policy specify the… | 1 | 1 | 0 | | 0 | 1.2.1 | 42 | 0 | 10200 |
| 10204 | Does it enforce that everyone us… | 1 | 1 | 0 | | 0 | 1.2.1 | 42 | 0 | 10200 |
| 10205 | Is there a process in place to d… | 1 | 1 | 0 | | 0 | 1.2.1 | 42 | 0 | 10200 |
| 10206 | Are there any safeguards in plac… | 1 | 1 | 0 | | 0 | 1.2.1 | 42 | 0 | 10200 |
| 10300 | Do you have a policy for creatin… | 2 | 1 | 0 | 1.3.1 | 1 | 1.3 | 7 | 0 | 0 |
| 10301 | Is there a policy for creating a… | 1 | 1 | 0 | | 0 | 1.3.1 | 14 | 0 | 10300 |
| 10302 | Do you have a process for tracki… | 1 | 1 | 0 | | 0 | 1.3.1 | 14 | 0 | 10300 |
| 10400 | Are you the only user with admin… | 3 | 1 | 0 | 1.4.1 | 1 | 1.4 | 7 | 1 | 0 |
| 10401 | How many administrator accounts … | 4 | 1 | 0 | | 0 | 1.4.1 | 49 | 0 | 10400 |
| 10402 | Is there a procedure that is fol… | 1 | 1 | 0 | | 0 | 1.4.1 | 49 | 0 | 10400 |
| 10403 | Is there a procedure or are ther… | 1 | 1 | 0 | | 0 | 1.4.1 | 49 | 0 | 10400 |
| 10404 | Are administrator accounts share… | 1 | 1 | 1 | | 0 | 1.4.1 | 49 | 0 | 10400 |
| 10405 | Are administartor accounts acces… | 1 | 1 | 1 | | 0 | 1.4.1 | 49 | 0 | 10400 |
| 10406 | Are the administrator accounts u… | 1 | 1 | 1 | | 0 | 1.4.1 | 49 | 0 | 10400 |
| 10407 | Are administrator passwords prot… | 1 | 1 | 0 | | 0 | 1.4.1 | 49 | 0 | 10400 |
| 10500 | Do you have an incident response… | 1 | 1 | 0 | | 0 | 1.5 | 7 | 0 | 0 |

*Figure 3: Questions table*

Non-persistent data includes user responses to questions, the result of calculations, and internal application data that feeds the application logic. Data classes were designed and instantiated to store this data. The use of data classes helps to streamline the flow of data and provides easier access to data throughout the application.

**4.2.2. Application design**

**4.2.2.1. Architecture**

In accordance with the Django web framework, the application consists of a number of files. Most importantly,

- views.py contains the core logic behind each page displayed to the user. When a page relies on user input, it distinguishes between GET and POST methods to handle data appropriately and avoid exposing sensitive data in the URL. It also has the branching logic depending on the user responses (from questions to sub-questions), database queries and calculations.

- models.py is where the database models are defined.

- urls.py stores the list of paths, linking each path to the corresponding function in views.py.

- admin.py holds the configuration for the admin backend of the application. It is possible to manipulate the databases from admin backend as long as they are registered here.

- helpers.py contains the data classes, instantiated objects, and several helper functions.

- dfd.py contains the code base to create and save the DFD.

- The templates folder stores the templates (.html files) used for all web

pages of the application. The Django template language is used to streamline the design and facilitate the flow of data throughout the application.

- The static folder hosts the images, stylesheets, and other assets.
- The media folder is where the DFD files are stored.

When the user visits a page, the corresponding function in view.py is called according to the URL configuration in url.py. The user's request is handled by the function according to the request type (whether GET or POST) and a response is displayed using templates.

**4.2.2.2. Application logic**

The questions in the application are organised into the six layers of the SEANCE framework and each layer is further broken down into steps to facilitate the branching logic of the application.

| Step | Question ID | Question | Parent? | Children |
|---|---|---|---|---|
| 1.1 | 010100 | Are you the sole owner of the business? (e.g., do you run the business alone or with some other co-owner or a partner?) | TRUE | 1.1.1 |
| 1.1.1 | 010101 | How many shareholders/partners do you have? | FALSE | None |
| 1.2 | 010200 | Do you have and enforce a password policy? | TRUE | 1.2.1 |
| 1.2.1 | 010201 | Does your password policy recommend that users create a different and unique password for each system they access? | FALSE | None |
| | 010202 | Does it provide guidance to users on how to create strong passwords? | FALSE | None |
| | 010203 | Does it specify the technical controls for measuring and managing the quality of passwords? | FALSE | None |
| | 010204 | Does it enforce that everyone uses their own account and that there are no shared accounts? | FALSE | None |
| | 010205 | Is there a process in place to detect and change the compromised passwords? | FALSE | None |
| | 010206 | Are there any safeguards in place to prevent passwords from being guessed by randomly trying all possible combinations (also known as "Brute Force")? | FALSE | None |
| 1.3 | 010300 | Do you have a policy for creating and deleting user accounts? | TRUE | 1.3.1 |
| 1.3.1 | 010301 | Is there a policy for creating a new user account? | FALSE | None |
| | 010302 | Are there any procedures in place to monitor user accounts as employees join and leave the organisation? | FALSE | None |
| 1.4 | 010400 | Are you the only user with administrator privileges (e.g., creating new users, setting passwords, configuring devices and network, etc.) in the business? | TRUE | 1.4.1 |

*Figure 4: Branching logic*

There are two categories of questions in the application: Parents with children (sub-questions) and non-parents without children. Each question can be one of the following five types: (1) those without children, (2) those that branch on a 'Yes' response, (3) those that branch on a 'No' response, (4) open-ended questions with no branching and (5) those that are only relevant to DFD (no value and no branching). Depending on the type of question and the user's answer, the main questions may lead to sub-questions in the next sub-step or to the next main question in the next main step.

With the exception of Type 5 questions that have no impact on the risk score, all questions carry equal weight in calculating the risk score for each layer and for the organisation as a whole. The cyber security risk score is calculated on a scale of 0 to 5, with the lower the score, the better the cyber security posture. Each Yes/No question is answered with either a "1" or "0" to be added to the tally if the answer is "Yes". A positive attitude, behaviour, or practice is scored as "0", reducing the total risk score, while a negative response is scored as "1". For instance, consider the following questions:

| Question | Value |
|---|---|
| Are administrator passwords protected by multi-factor authentication (MFA)? | 0 |
| Are the administrator accounts used for day-to-day activities? | 1 |

The first question represents a positive practice. So, if the user responds with 'Yes', no points will be added to the tally. However, if they respond with 'No', one point will be added to the total. Conversely, the second question reflects a negative attitude, with a value of one for 'Yes' and zero for 'No'. If the user selects 'I don't know', the question returns a value of 0.5. The value of the parent question is calculated by averaging the values of its children if they are activated. Otherwise, the system takes only its intrinsic value into account.

On the other hand, the values of open-ended questions need to be normalised before being added to the total. Therefore, their values are calculated by multiplying the user's input by 0.1. This is because the maximum number of employees in an MB is 10, and when multiplied by 0.1, it represents the maximum value of yes/no questions, which is '1'. For instance, if the user answers the question 'How many remote employees do you have, including yourself?' with '10', this reflects the highest level of risk for that particular question (10 x 0.1 = 1).

Each question also has a factor denominator that allows the values to be aggregated and the total score to be scaled between 0 and 1. As each question carries equal weight, this factor is effectively equal to the number of questions in a layer that contribute to the total score. In the case of parent questions, the factor denominator of sub-questions is multiplied by the total number of sub-questions of its parent in order to reflect the partial contribution of one sub-question to the total score. This factor normalises the impact of a question on a scale of 0 to 1. The value of each question is then multiplied by 5 to rescale from a 0-1 scale to a 0-5 scale.

In summary, after all questions in a layer have been answered, the risk score of the layer is calculated using the following formula:

$$\sum_{q=1}^{n} \frac{v_q}{f_q} * 5$$

where $q$ is the question, $n$ is the total number of questions, $v_q$ is the question value and $f_q$ is the factor denominator value of the question.

The overall risk score for the entire organisation is calculated by averaging the individual risk scores of each layer. Once the score is calculated, risk is assessed according to the table below:

| Score | Risk |
|---|---|
| 0 - 1.0 | Very low |
| 1.1 - 2.0 | Low |
| 2.1 – 3.0 | Moderate |
| 3.1 – 4.0 | High |
| 4.1 – 5.0 | Very high |

**4.2.2.3. Outputs**

Three outputs of the web applications are:

- **Risk score**: Using the formula above, a risk score is calculated on a scale of 0-5 for each layer and for the entire organisation, with higher scores indicating a greater threat.

- **Recommendations**: The application provides the user with recommendations based on threat indicators derived from the user's responses. The recommendations are based on industry-standard sources but are presented in an easy-to-understand, non-technical manner.

- **Data Flow Diagram (DFD)**: A DFD is generated from user input to

provide MB owners with a holistic view of their system, including their assets, employees and customers, and how they interact with each other.

Once the user has answered all the questions, all three outputs are presented to the user on the final screen and the user is provided with the option to download the resulting report in .pdf format.

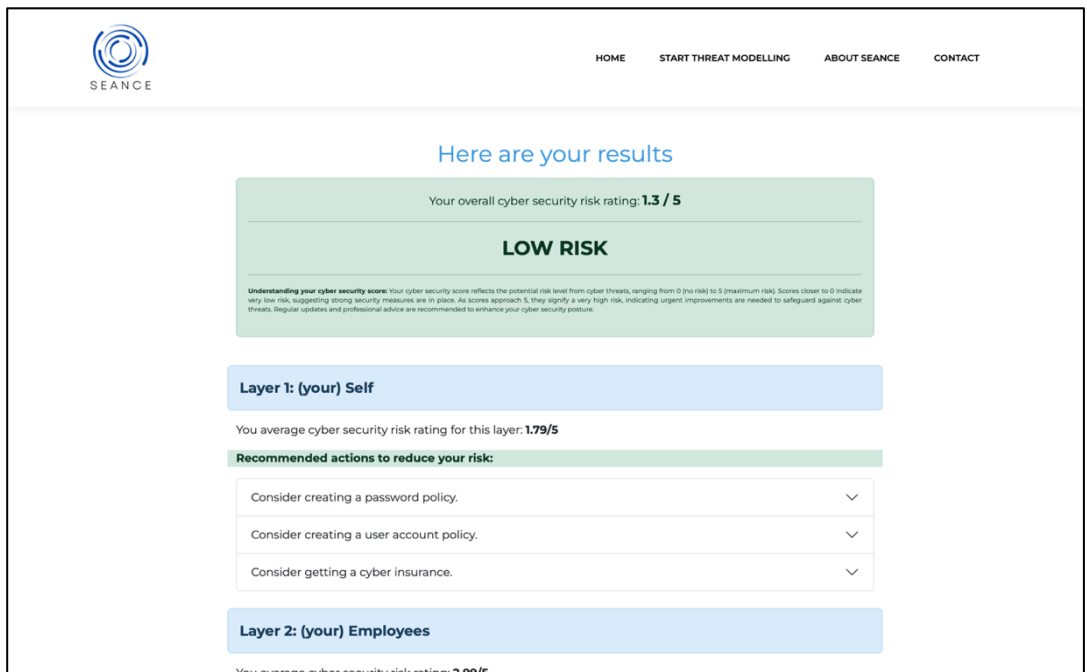

*Figure 5: Results page of the web application*

**4.2.2.4. Deployment**

The transition from the development environment to the production environment required a number of key changes to the application architecture:

- **Database**: For optimal performance, the database was migrated to PostgreSQL. For this purpose, the official PostgreSQL Docker image provided by Railway platform was used.

- **The storage and handling of static files and media files**: Storage for static and media files has been reconfigured so that both types of files are stored in separate folders ("/static" and "/media") in an AWS S3 bucket. This configuration reduces the load on the server, allows for the automatic deletion of files that are no longer needed, and scales storage as needed.

Finally, the application was deployed on the Railway platform. A Dockerfile was used to build the code to ensure that all Python and non-Python dependencies were installed and all necessary configurations (e.g. exposing the port and using gunicorn to launch the application) were made.

**4.2.2.5. Source Code**

The source code is available in the Github repository at https://bit.ly/seance-app

**5. Testing and Evaluation**

**5.1. Verification**

A total of 45 tests were written to verify the correct functioning of the application. Of these, 42 were unit tests and 3 were functional (API) tests, all of which were successfully passed.

```
1  Creating test database for alias 'default'...
2  Found 45 test(s).
3  System check identified no issues (0 silenced).
4  ----------------------------------------------------------------
5  Ran 45 tests in 0.396s
6
7  OK
8  Destroying test database for alias 'default'...
```

*Figure 6: Test results*

The tests were grouped into different categories and separate test files were created for models, views, helpers, DFD and API. For models and views, the test coverage was 100% and all key functions and classes were tested. The tests can be found in the 'tests' folder in the Github repository.

## 5.2. Validation

As discussed earlier in the literature review, despite the abundance of threat modelling and risk management frameworks currently available, it was found that all of them are to some extent technical, have a different focus or are not comprehensive enough to address all the needs of an MB owner. Accordingly, when evaluating the applicability of existing threat modelling frameworks to micro enterprises (RQ1), it was found that none of the existing frameworks met the requirements for a non-technical framework suitable for MB owners without cyber security expertise. However, some aspects and components of these frameworks were found to be applicable to MBs (RQ2) and these were explored in Section 3. In an attempt to design a non-technical framework, these findings have been incorporated into the design as much as possible to ensure that the newly designed framework and tool are aligned with industry standards, but at the same time can be used by people without cyber security expertise.

SEANCE Framework is based on three pillars: Layers, key themes, and guiding questions. The layered approach aims to provide MB owners with a way of thinking that will enable them to categorise and prioritise threats. The layers are designed to be easily relatable to an MB, such as the business owner themselves, customers, or the physical workplace (environment). The key themes in each layer highlight the most important cyber security concepts in that layer and help the MB owner to set the right focus. Finally, each layer has guiding questions that relate to the key themes. These simple questions are intended to assist MB owners to take the appropriate actions against potential cyber threats at each layer. The framework is supported with a web-based application based on user answering simple questions that are similar to the guiding questions of the framework. The application provides users with a risk score that helps MB owners assess their risk level at a glance. This is complemented by actionable recommendations to mitigate the risks identified. Finally, DFD provides a simple visual overview of the interaction between the business owner, employees, customers and business assets, indicating potential vulnerabilities.

The most distinctive feature of the framework and the web application is that it is free from any technical terminology and complex explanations that require technical knowledge. This makes the framework and application an ideal tool for MB owners who lack technical expertise, as evidenced by the results of the follow-up questionnaire in which the participants found the framework and the application easy to use, with only a few concepts and terms that they are not familiar with. In the light of these findings, the SEANCE Framework and the web-based threat modelling application are validated in relation to RQ3 as non-technical tools to help MB owners without cyber security expertise to enhance their cyber security posture.

### 5.2.1. Comparative analysis

This section compares SEANCE Framework and the web-based threat modelling tool with an industry-standard tool to highlight any potential differences in functionality, usability and effectiveness. The comparison will focus on the aim, theoretical background, scope and outputs of both instruments.

### 5.2.1.1. About Cyber Essentials Readiness Tool (CERT)

The Cyber Essentials scheme (UK Government, 2023), launched in 2014 as part of the UK's National Cyber Security Strategy, was intended to encourage companies to develop and sustain a certain degree of cyber resilience. It was designed as a non-mandatory certification scheme for businesses of all types. In this context, the CERT was introduced to assist organisations in identifying and implementing the minimum controls required to defend against cyber threats. Although not leading to the official certification, the CERT is structured as a questionnaire-based self-assessment tool, where the users are provided with an action plan based on their responses.

Given the similarities between the objectives and working methods of the SEANCE Framework and the CERT, it was deemed appropriate to conduct a comparison of the two in order to demonstrate the novelty and value of the SEANCE Framework.

### 5.2.1.2. Aim

CERT does not have a specific focus in terms of enterprise size; rather, it claims to appeal to businesses of all sizes. However, when the questions are examined, it is seen that CERT generally consists of non-technical questions and is limited to topics such as password and account protection, security of devices and basic network security. The only question which may be challenging for non-technical users is that about virtualisation. Therefore, it can be argued that the CERT does not meet the needs of large businesses, which may rely on complex IT infrastructures and have dedicated IT staff. As a result, the CERT appears to be more suited to small and micro-businesses and may be beneficial for businesses with simple infrastructure and no cyber security expertise.

Conversely, the SEANCE Framework and threat modelling tool are designed exclusively for MBs and do not address the requirements of larger businesses.

### 5.2.1.3. Scope

The CERT relies on a question base that is divided into 10 categories, in contrast with the six layers of the SEANCE Framework. However, the CERT categories appear to be more granular and not to follow any particular order, whereas the SEANCE Framework layers are more comprehensive and are organised in an inside-out approach.

A comparison of the themes that appear in the CERT with those in the layers of the SEANCE Framework reveals that almost all CERT themes align with those of the first four layers of the SEANCE Framework. However, the CERT does not address the fifth layer of the SEANCE Framework, which relates to customers, despite the significant risks associated with customer interactions with the business network and devices, as well as the privacy risks caused by the business' interactions with customers. It is particularly important to consider these risks in the context of the growing importance of privacy regulations and the potential legal and financial consequences of non-compliance. Additionally, Layer 6 also bears great importance as it addresses the risks associated with the environment in which the business operates. The security controls recommended in this layer range from firewall protection at the network boundaries, to physical controls such as cameras and fire safety measures. In this regard, the CERT recommends a firewall, but does not cover the physical security measures that would enhance the cyber security posture of a business.

In summary, it can be argued that CERT focuses on technical and administrative controls, whereas the SEANCE Framework suggests a comprehensive approach which includes technical, administrative and physical controls. Another significant distinction is that the SEANCE Framework considers non-organizational individuals, such as customers and other visitors, as part of the cyber security strategy, whereas the CERT does not.

| SEANCE Framework | | Cyber Essentials Readiness Toolkit | |
|---|---|---|---|
| Layer | Key themes | Categories | Themes |
| Layer 1: Self | Business owner(s) & Administrators<br>Organisation-wide security controls<br>Password policy<br>Incident Response Plan<br>Business Continuity Plan<br>Cyber Security Insurance | Use of passwords | Change default passwords<br>Password policy |
| | | User accounts | Administrator accounts<br>Account policy |
| Layer 2: Employees | Number of employees & Remote employees<br>User accounts & User training | Use of passwords | No shared passwords<br>Strong passwords |
| Layer 3: Assets | Computers (desktops, laptops, thin clients, terminals, etc.)<br>Servers<br>Cloud services<br>Web sites<br>Email accounts<br>IoT devices<br>Anti-virus/malware protection<br>Software/firmware updates<br>Data security<br>Data backups<br>Point of Sale (POS) System | Hardware or devices used by your organisation | Asset register<br>Thin clients<br>Servers |
| | | Software and firmware used by your organisation | Software/firmware list<br>Virtualisation<br>Automatic updates/Update policy<br>Unsupported software |
| | | Protection against malware | Anti-malware protection |
| | | Backing up data | Data backups |
| | | Secure configurations | Obsolete software<br>No default accounts<br>Disable AutoRun & AutoPlay<br>Locking mechanism for mobile devices |
| | | Cloud services | Shared Responsibility Model |
| Layer 4: Networks | Firewall<br>Routers<br>Internet access<br>Wi-Fi network<br>Cloud services<br>Shared Responsibility Model<br>Managed Service Provider (MSP)<br>Service Level Agreement (SLA)<br>Network monitoring | Firewalls | Firewalls on perimeter |
| | | Cloud services | List of cloud services<br>MFA<br>Shared Responsibility Model |
| | | Accessible services from the internet | Services accessible externally<br>Configuration over internet<br>Exposed services |
| Layer 5: Customers | Privacy<br>Wi-Fi Network | **NOT COVERED** | |
| Layer 6: Environment | Non-organisational people<br>Firewall<br>Physical security controls & Fire safety<br>Power outage<br>Theft | Firewalls | Firewalls on perimeter |

*Table 3: Comparison of SEANCE Framework and Cyber Essentials Readiness Toolkit*

#### 5.2.1.4. Theoretical Framework

In addition to its practical application, the SEANCE framework's primary objective is to raise awareness among MB owners about cyber security and encourage behavioural change. To achieve this, it has been designed as an easily memorable mnemonic and structured in layers. On each layer, key themes and guiding questions direct MB owners to potential areas of concern. The web-based threat modelling tool supports this framework by allowing MB owners to put the framework in practice. Conversely, the CERT approach does not have a theoretical basis and is solely based on practical implementation.

#### 5.2.1.5. Outputs

Both SEANCE and CERT provide users with actionable recommendations based on their responses. The reports generated by both tools provide valuable explanations of cyber security concepts and controls, along with suggestions on how to implement them. This would be beneficial for users without technical expertise. However, SEANCE offers two additional outputs that make it suitable and useful for non-technical users.

The first of these is the risk score which is calculated based on the user's responses on a scale of 0 to 5, with lower scores indicating a better cyber security posture. The risk score, which is a tangible and quantitative value, is a more straightforward and easily understandable indicator than text-based recommendations. This will, in turn, help a MB owner to appreciate their level of cyber resilience and raise their awareness. Furthermore, after implementing some security controls suggested by SEANCE, MB owners can revisit the tool later to see how these controls impacted their risk score. This allows them to track their cyber security efforts continuously.

Secondly, SEANCE provides a high-level DFD of the business, which demonstrates how the business owner, employees, customers, and assets interact with each other in a visual format, helping MB owners to identify potential vulnerabilities.

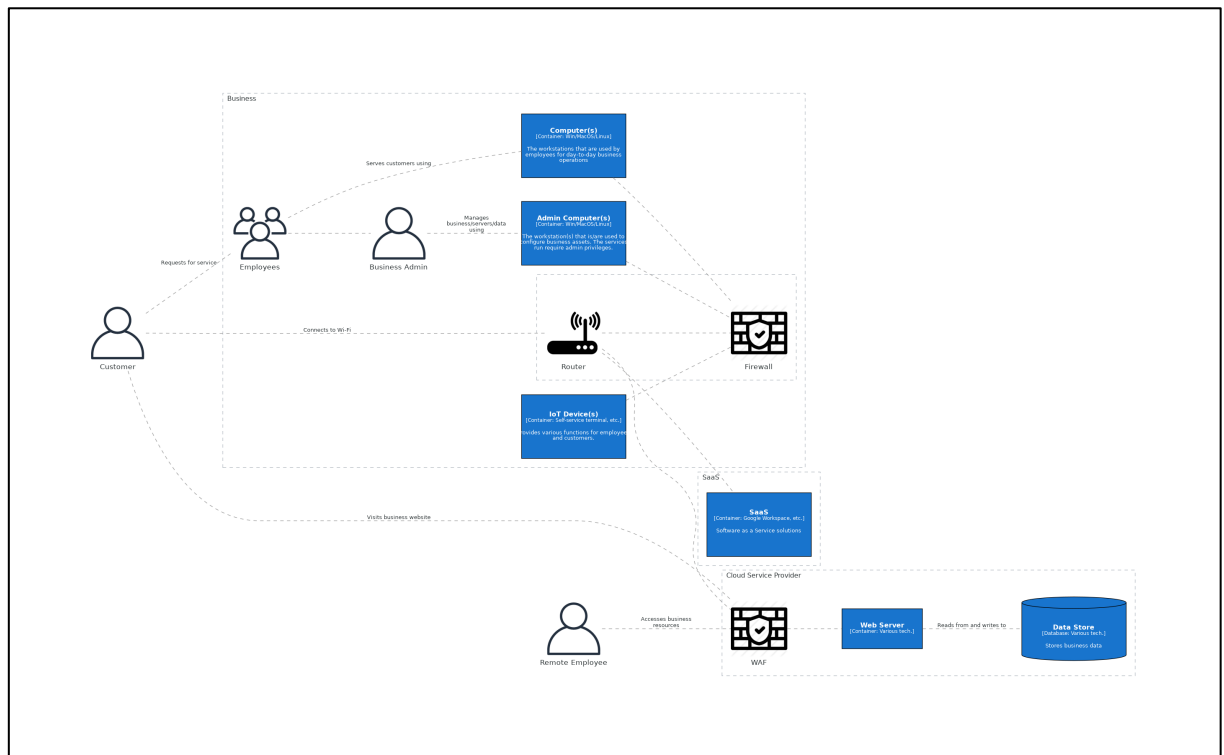

*Figure 7: A sample DFD produced by the SEANCE Threat Modelling Tool*

**5.2.1.6. Conclusion**

Although both tools provide valuable insights and recommendations to help MB owners to achieve higher levels of cyber resilience, as outlined above, there are significant differences between the two. CERT is designed as a starting point for a wider Cyber Essentials scheme backed by the UK Government, covering technical and administrative controls. On the other hand, SEANCE is a standalone, non-technical framework and a threat-modelling tool tailored for MBs looking to enhance their cyber security posture and it takes a more comprehensive approach, covering a broader range of measures ranging from technical to physical controls. While CERT only provides recommendations, SEANCE complements the recommendations it provides with a risk score and DFD, making the results more easily interpretable and understandable for the non-technical user.

## 5.3. Experiments and Analysis

### 5.3.1. Analysis of the initial survey results

Of the 320 MB owners and/or managers contacted, a total of 21 respondents consented to take part in the initial survey. After analysing the responses to the screening questions, businesses managed or owned by 19 respondents were found to meet the MB criteria. Two businesses were excluded from the analysis, one reporting a revenue of €4 million and the other reporting a total of 36 employees. Out of these, two respondents stated that they have a formal computer science and/or cyber security-related educational background, but no one indicated that they had a certificate in cyber security. This composition aligns well with the objective of the study as it aims to explore how MBs respond to cyber security threats in the absence of cyber security knowledge.

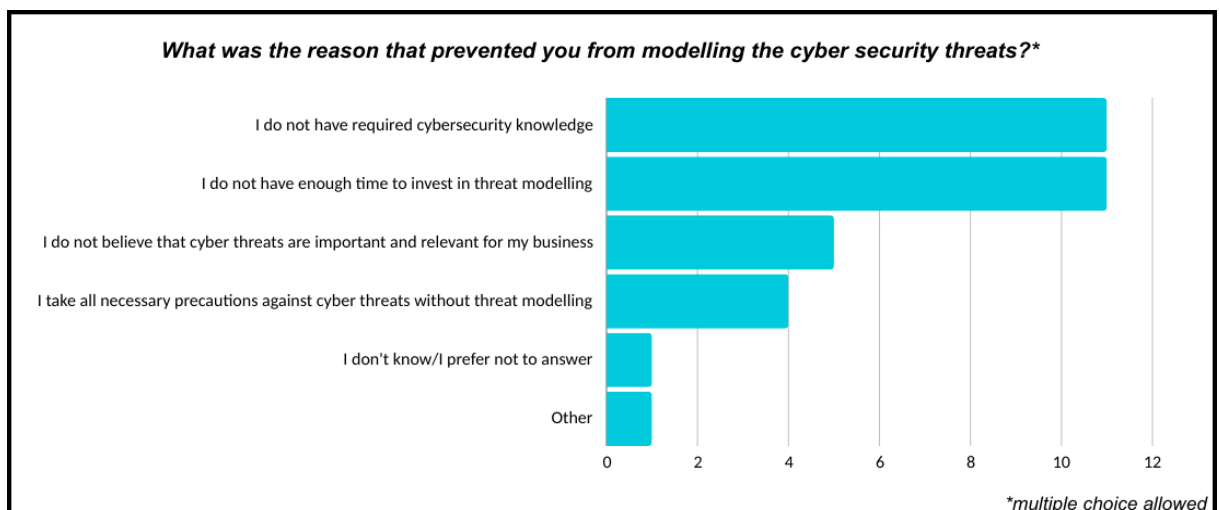

*Figure 8: Distribution of responses to the 'no' branch of question 5 of the initial questionnaire*

When asked if they had been engaged in threat modelling for their organisation, none of the respondents indicated that they had done so. The most common reasons for avoiding threat modelling were the lack of cyber security knowledge and the lack of time to invest in threat modelling. This was to be expected as MB owners usually do not have a background in computer science or cyber security, and the cyber security concepts are usually too complex for them to understand and apply. In addition, as mentioned earlier, MBs tend to have a small number of employees with everyone in the business having to perform many tasks at the same time, and many MBs are one-person organisations. In this environment, it is understandable that there is insufficient time to dedicate to an issue such as cyber security, which on the surface, does not directly contribute to the business. It is also worth noting that many MB owners think that they take all necessary cyber security measures. While this suggests that the cyber threats have been considered and addressed to some extent by the MB owner, it is difficult to gauge the effectiveness and longevity of the measures without a formal threat enumeration process. Finally, four respondents stated that they do not think cyber threats apply to their business, whereas it is not possible to think that businesses that have websites, many internet-connected devices at their workplaces and engage with their customers in cyber space are immune from cyber threats.

Consistent with the responses to the previous question, respondents pointed to the lack of time to invest in threat modelling and the lack of personnel with cyber security skills as the most significant challenges to threat modelling. Four respondents also indicated that they saw no challenge in modelling threats, potentially because of the limited knowledge about cyber security.

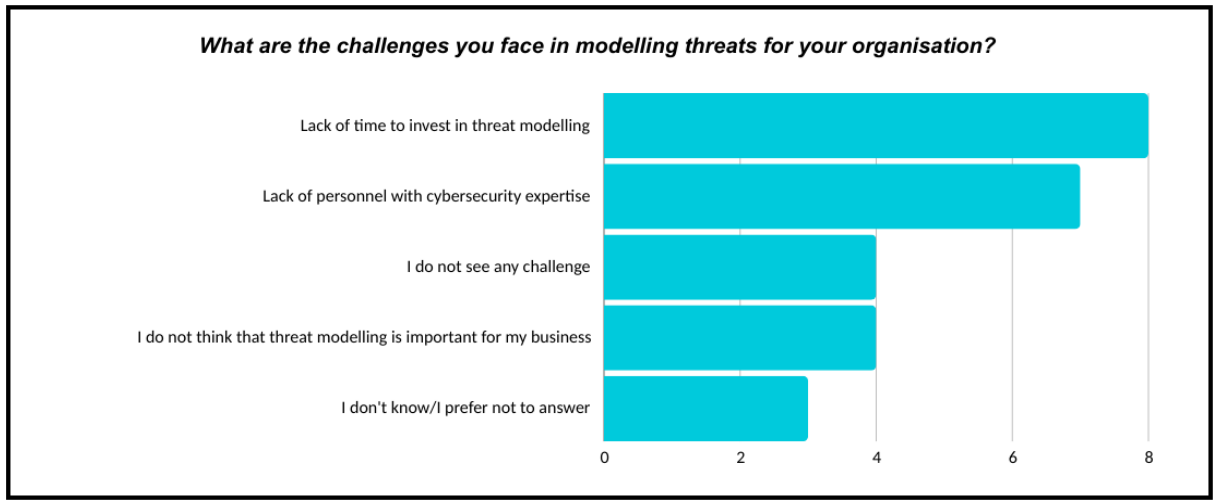

*Figure 9: Distribution of responses to question 8 of the initial questionnaire*

In conclusion, the survey results align with the findings of the literature review, particularly in terms of cyber security and threat modelling habits of MB owners and the difficulties they face in threat modelling.

**5.3.2. Analysis of the follow-up survey results**

A total of 14 participants, who had expressed their interest during the initial questionnaire, were invited to take part in the follow-up questionnaire. Of these, five participants experimented with the SEANCE Framework and threat modelling tool and responded to the questionnaire.

All participants confirmed that this was their first experience of reading about a threat modelling framework and using a threat modelling tool. All participants found the framework "fairly easy" to understand, which was the second-best option, and they all confirmed that they felt much more knowledgeable about threat modelling after reading about the framework. Only two participants stated that they encountered some concepts and terms that they were not familiar with. These were "MFA", "SaaS", "IoT" and "threat modelling".

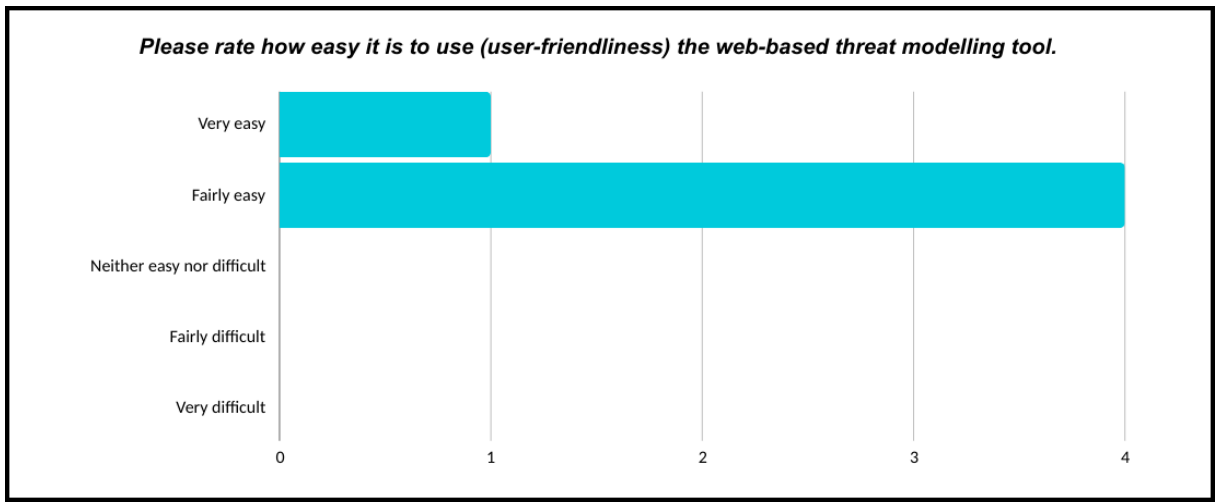

*Figure 10: Distribution of responses to question 6 of the follow-up questionnaire*

When asked to rate the ease of use of the threat modelling tool, the majority of respondents rated it as "fairly easy," with one respondent rating it as "very easy." All respondents stated that the tool helped them identify threats that they were previously unaware of. The threats mentioned by the respondents can be seen in Figure 9.

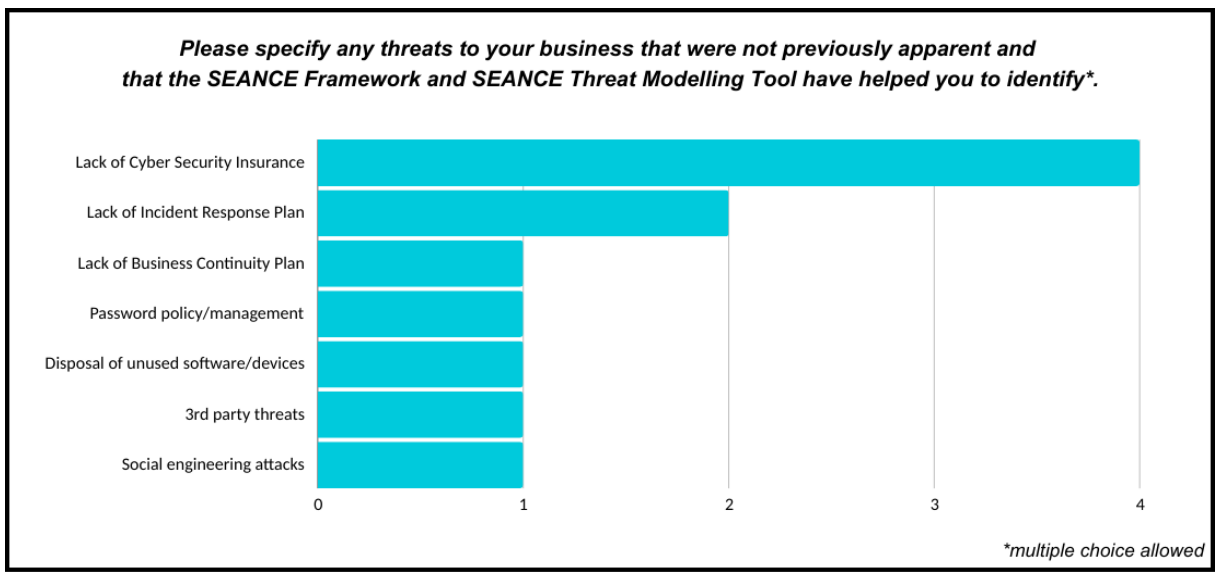

*Figure 11: Distribution of responses to the 'yes' branch of question 7 of the follow-up questionnaire*

Furthermore, after their experience with the SEANCE Framework and tool, all participants expressed their intention to utilise a similar framework and tool in the future to assist in the detection of cyber threats.

### 5.3.3. Discussion of the results

The results of the initial survey affirmed that MB owners generally do not have cyber security expertise and, largely for this reason, do not use a cyber security framework or tool. When contacted again after the development of the SEANCE Framework and web-based application and asked for their opinion, they confirmed that they found these tools useful and enumerated a number of threats that both artefacts helped them to identify. Consequently, when presented with a non-technical framework and tool, MB owners without cyber security expertise appreciated its value. This indicates that the SEANCE Framework and the SEANCE Threat Modelling Tool, or a similar non-technical framework tailored to non-technical stakeholders, could provide value and help reduce their cyber security risk. It is also noteworthy that all participants in the follow-up questionnaire indicated their intention to utilise similar tools for the identification of cyber threats, suggesting an enhanced awareness and a behavioural shift towards cyber security.

### 5.4. Lessons learned

In retrospect, several lessons can be drawn from the project that will help similar projects to be carried out more efficiently.

Questionnaires are indispensable tools for both qualitative and quantitative research. However, there are many factors that influence the response rate of surveys, such as length, medium, method of administration and number of people contacted, among others (Sue & Ritter, 2012; Deutskens et al., 2004). In this context, two important lessons have emerged from this project in relation to questionnaires. The first is that it is unrealistic to expect a response rate of more than 10 per cent. As this is quantitative research, the first survey in this project aimed for a minimum of 20 respondents, but 320 people had to be reached to achieve this number. Furthermore, although 14 participants in the first survey indicated that they wanted to participate in the second survey, only five actually did so. The second survey may have had a low participation rate as it involved familiarisation with the framework and experimenting with the tool rather than just answering the questions and required a significant time commitment. The second lesson is that more time should have been allocated to the survey. Although the project plan envisaged a 20-day period for sending out and receiving responses to the questionnaires, the low response rate and the need to reach more people to get more responses led to an extension of this period, delaying the overall project slightly. Therefore, by anticipating a lower response rate of around 5-10%, a more accurate calculation can be made of the number of people to be contacted based on the target number of responses, allowing for a more realistic time allocation for the surveys.

On a more technical level, deploying artefacts such as the web application presented in this project can be a daunting task, particularly due to the design choices that have been made. The transition from the development environment to the production environment can require significant refactoring due to the unique dependencies and requirements of each component used, as well as the technical debt incurred. It can also limit the range of deployment platforms and services available, as not all platforms provide the required functionality. For example, to draw the DFD, the SEANCE web application uses the *diagrams* Python package, which depends on the *pydot* and *graphviz* Python packages. However, for it to work properly, a system-wide installation of *graphviz* is also required. During development, the *diagrams* package was a logical choice due to its readily available features, but during deployment it proved to be too restrictive. Because *Graphviz* had to be installed, it was necessary to dockerise the application to deploy it (which would not have been necessary otherwise), further complicating the process and potentially increasing costs. Therefore, the minimisation of non-Python dependencies during the development may result in a more flexible, easier, and more cost-effective deployment process.

Testing is a crucial step in ensuring the correct functioning of software. The SEANCE web application was tested with 45 tests covering 100% of the views and models, and all tests passed. However, given the complexity of some of the functionality (e.g. the Questions view) and the unique characteristics of the Django web framework (i.e. the codebase is distributed across models, views and templates), Test-Driven Development (TDD) would have allowed for a more thorough and effective testing process. For example, all views were tested to ensure that they rendered the correct templates and returned the expected HTTP response codes, but the context and elements on the page were not tested because, as the functions were not written in a test-oriented way, this would have required significant refactoring of the tested functions, which was not possible due to time constraints. As shown by Percival (2017), the testing process would have been more robust if the TDD approach had been adopted by creating failing tests first and developing functions to satisfy and pass the tests later.

### 5.5. Future work

While the framework and the tool has been successfully validated against RQ3, both artefacts can be further improved with the introduction of some new functionalities or improving the existing features.

As demonstrated in Section 2 and Section 3, attack trees may provide the user with a holistic and hierarchical view of the threats that their business is susceptible to. However, during the development phase it was not possible to implement this feature without further complicating the application and user experience. It may be possible to generate an attack tree by capturing the parameters needed to draw an attack tree in a similar manner that the DFD parameters are captured from the user responses. In order to achieve this, it may be necessary to amend some of the existing questions or to add some new ones, the latter of which can be done simply by adding Type 5 questions to the database. Addition of this feature may prove useful especially in the prioritization of the measures that should be taken by the user.

In the current version of the web application, all questions and layers are equally weighted in the risk rating. However, it is evident that not all cyber threats have the same importance and impact. To address this shortcoming, the threat indicators can be more accurately weighted by using industry-standard resources like CVSS or OWASP Top 10, resulting in a more precise risk score. It is important to note that the application design enables such customisation with a simple adjustment of the factor denominator value attached to each question.

Techniques such as Persona non Grata and Security Cards, as mentioned in Section 3, can help an MB owner to appreciate the potential threats without the need for technical expertise. Integrating these techniques into the web application and/or the framework can significantly increase their value to the users.

## 6. Conclusion

The objective of the project was to examine and determine the suitability of the existing threat modelling frameworks for MBs, to investigate the challenges associated with cyber security practices in MBs and to develop a non-technical framework tailored to them.

MBs face a number of unique challenges when it comes to cyber security. In order to develop a better understanding of these challenges, a literature review and a survey was conducted among MB owners. As a result, lack of cyber security knowledge and lack of time to devote to cyber security were identified as the primary causes of these challenges, along with several secondary causes such as limited budget and shortage of manpower. It was also observed that there are some challenges preventing MBs from benefitting from the existing threat modelling and cyber security frameworks and tools. While there are many widely accepted threat modelling frameworks, each with its own focus and use case, none of the frameworks examined in this study were found to be fully suitable for MBs. Many of these are too complex to be used by non-technical MB owners, while others are too narrow in focus to cover the real-life activities of MBs. On the other hand, self-assessment tools such as the CERT were found to be very useful for MBs, but not comprehensive enough as they only cover technical and administrative measures. In light of these findings, the SEANCE Framework has been developed as a non-technical threat modelling framework tailored exclusively to meet the unique needs of MBs. Apart from being a non-technical framework, the novelty of the SEANCE Framework is that it incorporates customer interaction, privacy considerations and physical measures in cyber security assessments. Subsequently, a web-based tool that can be easily used by MB owners was developed based on this framework. The follow-up survey confirmed that both the framework and the tool helped MB owners to identify cyber security threats to their businesses.

Despite their limited size and turnover, MBs continue to have a large footprint in the economy (UK DBT, 2023; OECD, 2017). However, MBs remain a popular target for cyber criminals, with almost half of these organisations being targeted by cyber-attacks every year (UK DSIT, 2024). Nonetheless, the complexity of available tools prevents MB owners without a technical background from taking advantage of readily available cyber security know-how. Against this background, non-technical frameworks and tools, exemplified by the SEANCE Threat Modelling Framework and threat modelling tool developed in this project, can serve as an interface to help MB owners access the required information and improve the cyber security posture of their businesses. The assertion of this study is that such frameworks and tools need to be comprehensive to address technical, administrative and physical controls as well as emerging issues such as privacy, while maintaining ease of use and simplicity. In this project it has been demonstrated that it is possible to incorporate all these aspects into a non-technical framework and that it can be implemented in a user-friendly way, so that MB owners can put the framework into action to identify threats and implement necessary measures without technical expertise.

The first step in addressing the specific needs of MBs may be to differentiate between MBs and the other small and medium-sized enterprises (SMEs) rather than applying a one-size-fits-all approach under an overarching SME umbrella. An approach that treats these organisations as a homogeneous group may not yield the expected results, as MBs may lack the financial or human resources to invest in cyber security, while the other SMEs are generally more comfortable in this regard.

This paper outlines a customised cyber security solution for MBs. Similar projects, with a specific focus on MBs, designed and implemented with the support of governmental and recognised non-governmental actors, may help MBs to become more cyber-resilient.

IBM Security (2023) Cost of a Data Breach Report 2023. Available from: https://www.ibm.com/downloads/cas/E3G5JMBP [Accessed 12 February 2024].

ISO (2022) ISO/IEC 27002:2022 - Information security, cybersecurity and privacy protection — Information security controls. Available from: https://www.iso.org/standard/75652.html [Accessed: 12 February 2024].

Jamil, H. (2022) *Factors Affecting Users Cybersecurity Practices: A Study of Australian Microbusinesses*. Ph.D. thesis, Charles Sturt University. Available from: https://researchoutput.csu.edu.au/en/publications/factors-affecting-users-cybersecurity-practices-a-study-of-austra [Accessed: 12 February 2024].

Kabanda, S., Tanner, M. & Kent, C. (2018) Exploring SME cybersecurity practices in developing countries. *Journal of Organizational Computing and Electronic Commerce* 28(3): 269–282. DOI: https://doi.org/10.1080/10919392.2018.1484598

Kohnfelder, L. & Garg, P. (1999) The threats to our products. Available from: https://www.first.org/global/sigs/cti/curriculum/The-Threats-To-Our-Products.docx [Accessed: 12 February 2024].

LeBlanc, D. (2007) DREADful. Available from: https://learn.microsoft.com/en-us/archive/blogs/david_leblanc/dreadful [Accessed: 28 January 2024].

Mangaoang, E., Monreal, R. N., Mangaoang, E. F. & Monreal, R. N. (2024) Common Vulnerabilities and Exposures Assessment of Private Higher Educational Institutions Using Web Application Security. *J. Electrical Systems* 20(5): 668–676. DOI: https://doi.org/10.52783/jes.2288

Mead, N., Shull, F., Spears, J., Heibl, S., Weber, S. & Cleland-Huang, J. (2017) Crowd Sourcing the Creation of Personae Non Gratae for Requirements-Phase Threat Modeling. *Proceedings - 2017 IEEE 25th International Requirements Engineering*

Survey 2024. Available from: https://www.gov.uk/government/statistics/cyber-security-breaches-survey-2024/cyber-security-breaches-survey-2024 [Accessed: 16 May 2024].

UK Government (2023) Cyber Essentials scheme: Overview. Available from: https://www.gov.uk/government/publications/cyber-essentials-scheme-overview [Accessed: 18 May 2024].

Wilson, M., McDonald, S., Button, D. & McGarry, K. (2023) It Won't Happen to Me: Surveying SME Attitudes to Cyber-security. *Journal of Computer Information Systems* 63(2): 397–409. DOI: https://doi.org/10.1080/08874417.2022.2067791

Xiong, W., Legrand, E., Åberg, O. & Lagerström, R. (2022) Cyber security threat modeling based on the MITRE Enterprise ATT&CK Matrix. *Software and Systems Modeling* 21(1): 157–177. DOI: https://doi.org/10.1007/s10270-021-00898-7

Yudhiyati, R., Putritama, A. & Rahmawati, D. (2021) What small businesses in developing country think of cybersecurity risks in the digital age: Indonesian case. *Journal of Information, Communication and Ethics in Society* 19(4): 446–462. DOI: https://doi.org/10.1108/JICES-03-2021-0035

## Appendix A: Initial Questionnaire

**1. How many people are working for your business?**

    ____

    ☐ I prefer not to answer

**2. What is the approximate annual turnover of your business?** This data will only be used to determine the size of your business (i.e. micro, small, medium, etc.). Please indicate the currency (USD, EUR, etc.).

    __________________

    ☐ I prefer not to answer

**3. Do you have a formal education in Computer Science, Cybersecurity or a related field?**

    ☐ Yes

    ☐ No

    ☐ I prefer not to answer

**If yes, what is the highest level of education you have in the fields mentioned above?**

    ☐ BSc

    ☐ MSc

    ☐ PhD

    ☐ Other (Please specify: ______________________ )

    ☐ I prefer not to answer

**4. Do you hold a professional certification (e.g. CISSP, ComTIA Security+, etc.) in cyber security? If yes, please indicate.**

☐ Yes

☐ No

☐ I prefer not to answer

**If yes**, please specify the certificate(s) you currently hold:

*Before moving on to the next question, please read this informative piece about threat modelling.*

Threat modelling is an activity of collecting and organising all information about possible threats to the valuable assets of a system, network, business process or application. This activity results in a threat model that structurally represents all information affecting the security of those assets. Threat models are presented in various forms, the most popular being threat diagrams and attack trees. Below is a sample attack tree (taken from *Secrets and Lies* by Bruce Schneier):

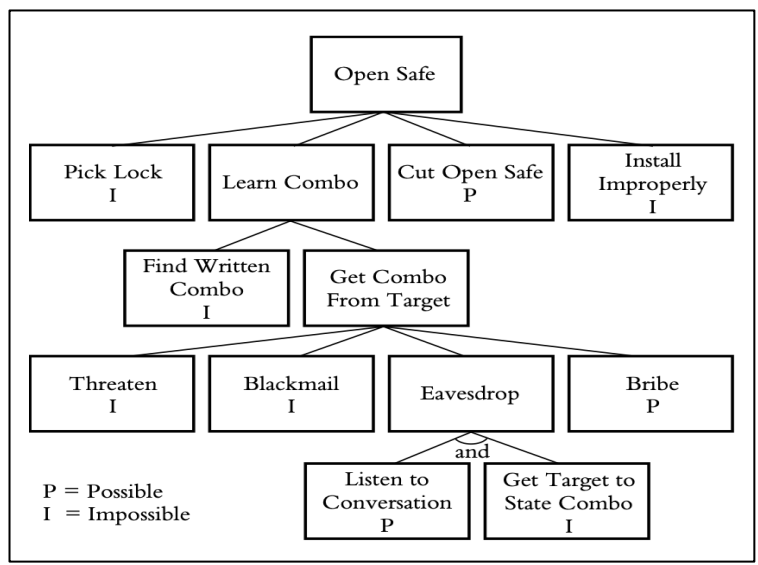

*An example of an Attack Tree*

**5. Have you ever modelled the cybersecurity threats against your business?**

☐ Yes

☐ No

☐ I prefer not to answer

**If yes, do you have an up-to-date threat model for your business at the moment?**

☐ Yes

☐ No

☐ I prefer not to answer

**If no, what was the reason that prevented you from modelling the cybersecurity threats?** Please select all that apply.

☐ I do not believe that cyber threats are important and relevant for my business.

☐ I do not have required cybersecurity knowledge.

☐ I do not have enough time to invest in threat modelling.

☐ I take all necessary precautions against cyber threats without threat modelling.

☐ I do not know/I prefer not to answer.

☐ Other (Please specify: _______________________________)

**6. Please list the threat modelling frameworks that you have heard about before or that you are familiar with:**

**7. Have you ever used a threat modelling tool (e.g. Microsoft Threat Modeling Tool, OWASP Threat Dragon, Cairis, etc.) before?**

☐ Yes

☐ No

☐ I prefer not to answer

**If yes**, please specify the tool(s) that you have used:

**8. Please indicate the challenges you face in modelling threats for your organisation (select all that apply):**

☐ I do not see any challenge

☐ Lack of personnel with cybersecurity expertise

☐ Lack of time to invest in threat modelling

☐ I do not think that threat modelling is important for my business

☐ I do not know/I prefer not to answer

☐ Other (Please specify: _______________________________)

**9. Are you willing to participate in a follow-up questionnaire that will be conducted within a few months?** This questionnaire will possibly involve reading briefly about a threat modelling framework, using a threat modelling tool and responding to a few questions. If you decide not to take part in the second questionnaire, you will not be contacted again in connection with this study.

☐ Yes

☐ No

# Appendix B: Follow-up Questionnaire

*Now, please take some time to read and familiarise yourself about the SEANCE Framework.*

The SEANCE Framework is a threat modelling framework developed by Etkin Getir (researcher) to help micro business owners and/or managers identify and mitigate cyber security risks to which their businesses may be susceptible.

In a nutshell, the SEANCE Framework provides for threats to be considered and assessed in six layers, with the general assumption that the most dangerous threat is the closest threat. Therefore, when considering threats to your business, you should think about **your:**

- **S**elf
- **E**mployees
- **A**ssets
- **N**etwork
- **C**ustomers
- **E**nvironment

**Before proceeding to the questions,** please read about the framework [here](#).

**1. Was this the first time you have ever read about a threat modelling framework?**

☐ Yes

☐ I am not sure/I prefer not to answer

☐ No

**2. Do you feel yourself more knowledgeable about threat modelling after reading the threat modelling framework brief?**

☐ Yes

☐ I don't know/I prefer not to answer

☐ No

**3. How difficult was it to understand the threat modelling framework brief presented to you?**

☐ Very easy

☐ Fairly easy

☐ Neither easy nor difficult

☐ Fairly difficult

☐ Very difficult

☐ I prefer not to answer

**4. Have you come across any concepts/terms that are unfamiliar to you when reading the threat modelling framework? If so, please specify.**

☐ Yes

☐ No

☐ I prefer not to answer

**If yes, please specify the the terms/concepts that you are unfamiliar with:**

***Now, please take some time to experiment with the SEANCE Threat Modelling Tool.***

The SEANCE Threat Modelling Tool is a web-based threat modelling tool developed based on the SEANCE Framework. It is designed to be as simple as possible, taking into account that the users of the tool (micro business owners and/or managers) are generally not experts in cyber security.

After asking the user a few simple questions, the tool calculates a risk score based on the user's answers to the questions, provides recommendations and draws a data flow diagram of the business. Users can download the report in PDF format if they wish.

*SEANCE Threat Modelling Tool Results Page*

**Before proceeding to the next set of questions,** please experiment the SEANCE Threat Modelling Tool by clicking here.

**Please experiment the tool on a <u>desktop or laptop computer</u> for the best experience, as has not been optimized for mobile devices yet.**

We kindly ask you to **answer all questions that will be posed to you and reach the results page on which you will be presented your risk score, recommendations for your business and a data flow diagram**. <u>Please note that you can experiment the tool as much as you like</u> (e.g. start from scratch, give different answers to the questions and see their impact) as we neither collect any results or interested in the results (i.e. risk score, data flow diagram, etc.) for the sake of this study.

**5. Was this the first time you have ever used a threat modelling tool?**
- [ ] Yes
- [ ] I am not sure/I prefer not to answer
- [ ] No

**6. Please rate how easy it is to use (user-friendliness) the web-based threat modelling tool:**
- [ ] Very easy
- [ ] Fairly easy
- [ ] Neither easy nor difficult
- [ ] Fairly difficult
- [ ] Very difficult

**7. Have the threat modelling framework and web-based threat modelling tool helped you identify a threat/threats to your business that were not previously apparent?**

- [ ] Yes
- [ ] No
- [ ] I prefer not to answer

**If yes, please specify the the any threats to your business that were not previously apparent and that the SEANCE Framework and SEANCE Threat Modelling Tool have helped you to identify.**

**8. After your experience with the SEANCE Framework and Threat Modeling Tool, do you intend to use a threat modelling framework or tool in the future (remember that this tool was developed as a result of a research project and may not be available after the project)?**

- [ ] Yes
- [ ] I don't know/I prefer not to answer
- [ ] No

**9. Do you have anything you'd like to add?**

# Appendix C: SEANCE Threat Modelling Framework

SEANCE is an innovative, non-technical threat modelling framework tailored for micro businesses. In addition to improving the overall level of cybersecurity in the business, it also aims to increase cybersecurity awareness and create behavioural change among micro business owners.

Based on the idea that threats can be abstracted in layers from the inside out, the framework consists of six layers. The fundamental logic is that a weakness in the inner layer will propagate to other layers; therefore, weaknesses in the inner layers are prioritised over the outer layers. For example, if no password policy is adopted by the owner of the micro business (Layer 1), the likelihood of strong passwords being used by employees (Layer 2) is significantly diminished.

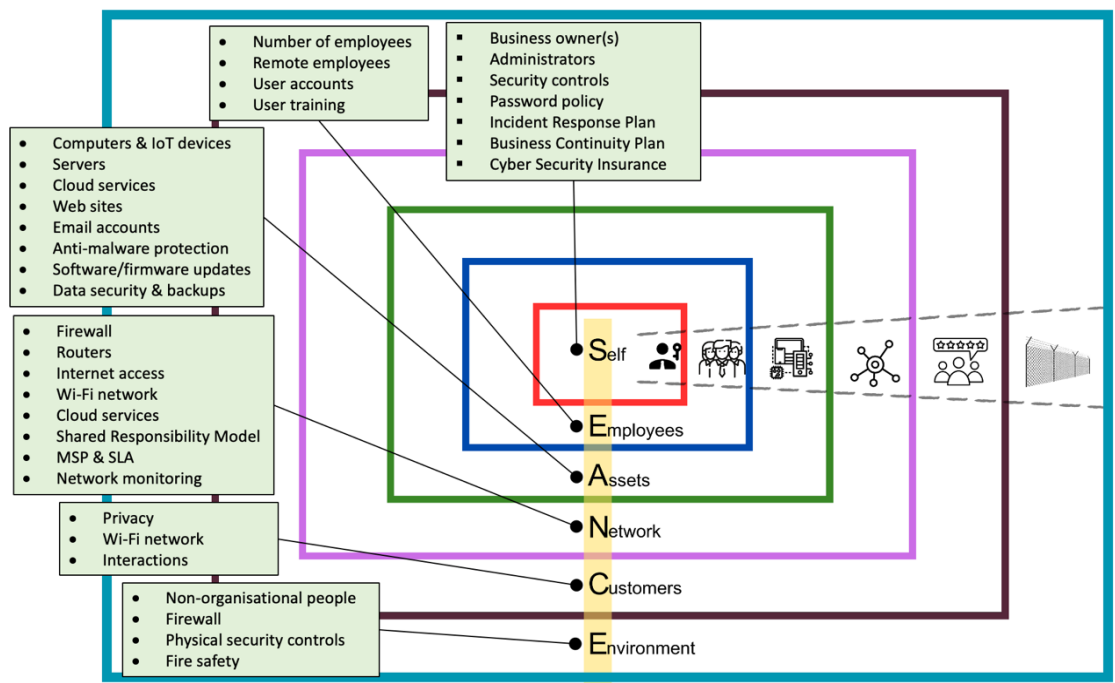

*Seance Framework and key themes at each layer*

The six layers of the SEANCE Framework are:

1. (your)Self
2. (your) Employees

3. your (Assets)
4. (your) Network
5. (your) Customers
6. (your) Environment

Below is a detailed description of each layer and the design decisions taken during the design of the framework. Key themes and key guiding questions for each layer are also provided. The factsheet that was presented to participants in the follow-up questionnaire and published on the 'About SEANCE' page of the SEANCE Framework web application is provided in Appendix D, along with the full set of guiding questions.

**Layer 1: Self**

The innermost layer of the framework is where the threats to the MB owner, or in the case of multiple ownership, the MB owners, and, more importantly, the security controls put in place and managed by the MB owner are reviewed and assessed. Ranging from password policies to cyber security insurance, these considerations are of the utmost importance because any shortcomings or negligence at this layer can have implications on the outer layers. On the other hand, the security controls and measures at this layer are the easiest for the MB owner to implement and configure as they have direct control.

To illustrate the importance of security controls at this layer, let us consider a password policy. If the password policy fails to cover all possible scenarios and is not aligned with best practices, enforcing it will not mitigate password-related risks, even if all employees fully comply.

| Key themes | Key guiding questions |
|---|---|
| • Business owner(s)<br>• Administrators<br>• Organisation-wide security controls<br>• Password policy<br>• Incident Response Plan<br>• Business Continuity Plan<br>• Cyber Security Insurance | • Are you the sole owner of the business or are there other shareholders/partners who run the business with you?<br>• Do you carry out the administrative tasks yourself, or do you have a designated admin?<br>• Do you have and enforce a password policy?<br>• Do you have a policy for creating and deleting user accounts?<br>• Do you have a policy for managing administrator accounts?<br>• Do you have an incident response plan in place in the event of a cyber security incident?<br>• Do you have a business continuity plan to get your business back up and running as quickly as possible?<br>• Do you have a cyber security insurance in place to cover or reduce the financial risks caused by cyber security events? |

**Layer 2: Employees**

The second layer comprises the staff who work for the MB. Employees may be full-time, part-time, volunteer, temporary or seasonal. It is important to assess the level of access and permissions given to employees within the organisation, as well as their understanding of security protocols and procedures. In addition, regular training and awareness programmes can help ensure that all employees are equipped to deal effectively with potential security threats. As remote workers bring a whole new level of complexity to cyber security in the post-pandemic world, it is also important to consider their unique characteristics. It is also worth noting that the number of employees has a significant impact on the cyber security posture of the MB: while fewer employees can lead to multi-tasking and shared accounts, more employees can sometimes lead to less control but better user management.

The second layer brings all of these considerations together to help MB owners better prepare their employees for potential cyber threats and to anticipate and prevent the consequences of employee misconduct.

| Key themes | Key guiding questions |
|---|---|
| <ul><li>Number of employees</li><li>Remote employees</li><li>User accounts</li><li>User training</li></ul> | <ul><li>How many employees do you have?</li><li>Do you have any remote employees who need to access to company assets remotely?</li><li>Do you regularly brief and train all employees on the procedures and processes related to the organisation's cyber security measures and policies?</li><li>Do you have a password policy for the users?</li></ul> |

**Layer 3: Assets**

Layer 3 focuses on the assets of the organisation, including physical assets such as various equipment used in the workplace, such as computers or self-service terminals, and intangible assets such as data and software. This layer encourages MB owners to think about and review the controls in place to secure their assets and how they can be improved.

This layer consider not only computers in the traditional sense, but also IoT devices that are growing in popularity and use. By considering all types of assets, MB owners can better protect their businesses from potential threats, especially those posed by IoT devices that are seemingly innocuous and straightforward. When it comes to device security, the importance of software and firmware updates cannot be overstated. The "don't break what works" attitude of the average user and some MB owners can be detrimental to their business, as these updates are the only way that developers can fix discovered vulnerabilities. While the NCSC recommends that the latest updates (2021) are installed automatically, if for some reason (e.g. to avoid unexpected downtime during updates) automatic updates are not preferred, the software and firmware should be patched as soon as possible. In this respect, it is also important to dispose of devices that are no longer supported and no longer receive updates, or to disconnect them from the Internet. Anti-virus and anti-malware protection, which MB owners should already be familiar with, completes the device security measures that should be taken at this layer.

Data is arguably the most valuable asset an organisation has. Data security measures should be considered in two categories: To minimise data loss in the event of a cybersecurity event, the business should have procedures in place to back up its data, which should include recovery point objective (RPO) and recovery time objective (RTO). The other aspect of data security, usually avoided by MB owners due to its inherent complexity, is the encryption of data, both at rest and in transit, which would help MB owners reduce the likelihood of sensitive information being exposed. While these concepts may be difficult for MB owners to understand, the SEANCE Framework encourages MB owners to consider these issues with questions that have been rephrased in an easily understandable manner. Using the services of a cloud service provider or managed service provider (MSP) may change how data backups and encryption are handled, which should be part of service level agreements (SLAs), as discussed at the next layer.

| Key themes | Key guiding questions |
|---|---|
| - Computers (desktops, laptops, thin clients, terminals, etc.)<br>- Servers<br>- Cloud services<br>- Web sites<br>- Email accounts<br>- IoT devices<br>- Anti-virus/malware protection<br>- Software/firmware updates<br>- Data security<br>- Data backups | - How many computers do you use to run your business? Consider all the desktop and laptop computers that you use to store data, manage orders, and communicate with the customers.<br>- Do you keep an inventory of your assets?<br>- Do you operate your own servers?<br>- Do you use a cloud service provider to store data, run a business application, etc.?<br>- Do you use Software as a Service (SaaS) products such as Google Docs, Microsoft365, etc.?<br>- Do you use IoT devices?<br>- Have you enabled automatic updates for all of your applications, including the operating system (OS)?<br>- Are you using software that is no longer supported?<br>- Do you review your devices regularly and disable or remove unused software?<br>- Are you making sure that all accounts on your devices and cloud services are only those that are used as part of your day-to-day business?<br>- Have you disabled Auto-Run and Auto-Play on all your devices?<br>- Do you lock your mobile devices with a locking mechanism like a password, fingerprint, facial recognition, etc.?<br>- Do you ensure that all default passwords on all devices are changed? |

|  | - Are all your computers, laptops and mobile devices protected against malware?
- Do you restrict the installation of third-party (unsigned) applications outside the official application stores?
- Do you keep your sensitive and critical data encrypted when it is stored (at rest) or while it is being transmitted (in transit)?
- Do you securely delete and/or destroy the data that is no longer in use?
- Do you have a procedure in place for regularly backing up your data?
- Do you have a website?
- Do you have procedures in place to safely dispose of and destroy devices (i.e. hard drives, flash drives, etc.) that are no longer in use? |
|---|---|

**Layer 4: Network**

As the name suggests, this layer is all about the organisation's network and how it is connected to the outside world, i.e. the Internet. An important aspect of this layer is cloud services, the use of which can dramatically change security considerations. This is because cloud services can come in different flavours such as SaaS, PaaS, IaaS, each with different responsibilities for the MB owner. Therefore, the network layer encourages MB owners to review the Shared Responsibility Model to understand what to expect from the cloud service provider in terms of cyber security and what falls on their shoulders. It is also important to review the SLA with the cloud service provider to ensure it meets the needs of the business.

Firewalls and routers play a central role in this layer as they are the main devices that help MB owners to manage, filter and secure the traffic on their networks. Firewalls at this level are used to separate the corporate network and create a demilitarised zone (DMZ). On the other hand, routers with firewall capabilities should be sufficient for most MBs, eliminating the need to invest in and manage a separate firewall. However, as it is highly likely that MB owners do not have the necessary knowledge and skills to properly manage their networks, they can seek help from an MSP to fill this gap. However, the MB owner should ensure that the expectations of the MSP are captured in an appropriate SLA. MB owners may not be aware of the importance of this agreement, and this layer aims to address this shortcoming by encouraging MB owners to learn more about SLAs and to review their SLAs (if they have any) or carefully prepare one when entering into an agreement with an MSP.

Finally, if we consider a micro-business such as a coffee shop or bookshop, while the Wi-Fi network in the workplace and the access it provides to customers may seem well-intentioned and harmless, it can act as a backdoor for malicious users to infiltrate the business network and assets or the devices of the business owner or employees (as well as those of other customers). Therefore, proper configuration of the Wi-Fi network, including the use of firewalls, is crucial.

| Key themes | Key guiding questions |
|---|---|
| • Firewall<br>• Routers<br>• Internet access<br>• Wi-Fi network<br>• Cloud services<br>• Shared Responsibility Model | • Is your business network protected by a firewall (or a router with firewall capability)?<br>• Do you have any services enabled that can be accessed externally?<br>• Can your routers or hardware firewalls be configured remotely over the Internet? |

| - Managed Service Provider (MSP)<br>- Service Level Agreement (SLA)<br>- Network monitoring | - Have you configured your routers or your hardware firewalls to block all other services being advertised to the Internet?<br>- Do you keep the firmware on your router(s) updated?<br>- Do you use the services of a third-party Managed Service Provider (MSP) to manage your network and network devices and to monitor the devices and networks for suspicious activity?<br>- Do you use cloud services in your organisation?<br>- Do you allow people outside your organisation (e.g. customers, suppliers, etc.) to connect to the same Wi-Fi network that your business assets are connected to? |
|---|---|

**Layer 5: Customers**

This layer is the thinnest in the framework, as the MB owners do not have direct control over customers. However, the behaviour of customers has a significant impact on the organisation's cyber security posture as they interact with the organisation's assets such as the website, ordering systems and so on. The framework distinguishes customers from other non-organisational people because they are allowed to penetrate the external boundaries of the organisation in a structured way to bring in revenue, which is important for the longevity of the business, and there is assumed to be a constant flow of customers as opposed to other non-organisational people (e.g. government officials, suppliers, etc.) who have a sporadic flow.

In recent years, privacy has gained prominence due to increased concerns about data breaches and unauthorised access to personal information. Privacy regulations have also become more stringent to protect individuals' data and ensure that companies are held accountable for any breaches. This has led to a greater emphasis on data security measures and transparency in how personal information is collected and used. Privacy and data protection measures must be implemented to ensure that customer information is protected from cyber threats. This is also a critical step in protecting the business from potential liabilities that would result from non-compliance.

| Key themes | Key guiding questions |
|---|---|
| - Privacy<br>- Wi-Fi network | - Do you collect personally identifiable information (PII) from customers for any purpose (marketing, research, etc.)?<br>- Do you allow users to create and maintain accounts on your site?<br>- Do you allow customers to connect to the same Wi-Fi network that your business assets are connected to?<br>- How do customers interact with your assets, network, employees, and yourself? |

**Layer 6: Environment**

The outermost layer can be thought of as the external boundaries of the organisation, which almost always overlap with its physical perimeter. Most of the measures that should be taken at this layer are physical security controls that are less technical in nature, such as security cameras, locks, guards, etc.

The firewall, the only proposed technical control at this layer, reappears here.

Unlike the internal firewall at the network layer, its purpose here is to protect the organisation from malicious traffic from outside the organisation. In a defence-in-depth approach, this will act as the first layer of protection until the filtered traffic is received by the internal firewalls.

| **Key themes** | **Key guiding questions** |
|---|---|
| - Non-organisational people<br>- Firewall<br>- Physical security controls<br>- Fire safety<br>- Power outage<br>- Theft | - Do you have a firewall enabled to filter/block incoming external traffic?<br>- Approximately how many people from outside your organisation, other than customers, visit your workplace (per day/week/month/year)? Think of suppliers, government officials, etc.<br>- Do you have physical security controls (e.g. cameras, locks, etc.) in place to secure the external perimeter of the organisation?<br>- Do you have the necessary fire safety measures in place?<br>- Do you have uninterruptible power supplies (UPS) or similar in place to keep critical equipment running and prevent data loss in the event of a power outage? |

**Appendix D: SEANCE Threat Modelling Framework Factsheet**

SEANCE is a threat modelling framework tailored for micro businesses (i.e., businesses with fewer than 10 employees and a turnover of no more than €2 million). It is a non-technical, asset-centric framework that aims to be easily understood and used by micro business (MB) owners who do not have the cybersecurity expertise or resources (i.e. financial and otherwise) to use an existing framework, tool, or system to identify the threats to their business.

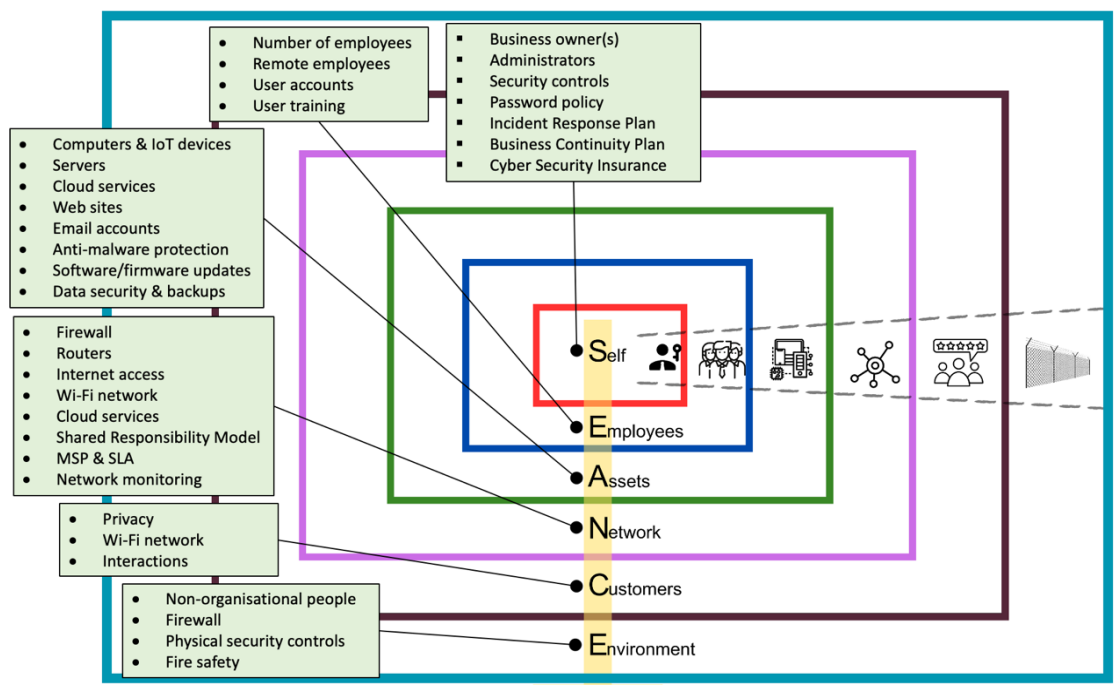

The SEANCE framework provides for threats to be considered and assessed in six layers, with the general assumption that the most dangerous threat is the closest threat. Therefore, when considering threats to your business, you should think about your:

- **S**elf
- **E**mployees
- **A**ssets
- **N**etwork

- **C**ustomers

- **E**nvironment

You can think of your organisation as a medieval castle: The lord (or commander for that matter) of the castle makes all the important decisions **himself** that have an impact on the whole castle. His 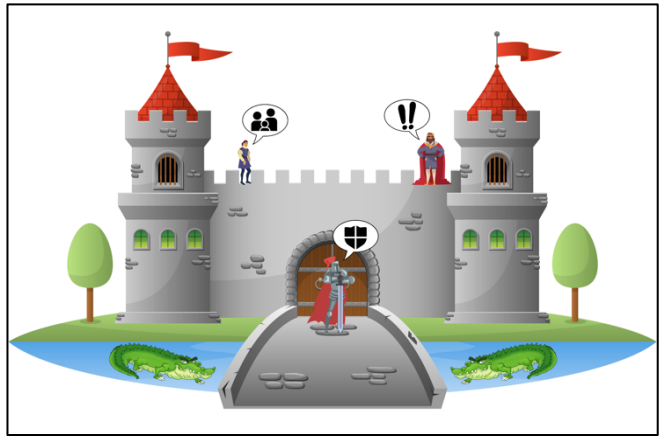 army of warriors (in local jargon, his **employees**) carry out his decisions by using their weapons and other **assets** in accordance with his wishes. This is not an individual endeavour, however; it requires a castle-wide **network** to coordinate the deployment of weapons and smaller units, with the ultimate goal of keeping the castle's defensive posture in sync. Then there are the people who live within the castle's perimeter, who are usually unaware of their lord's important defensive decisions, and who are more concerned with the livelihood of their household, paying their lord to be exempted from military service - perhaps that's why the lord mocks them by referring to them as his **customers** at his festive dinners (where he may consume too much beer). The castle's **environment** then has a huge impact on the castle's security: Unlike a peaceful castle on a remote island, about which the local population hears (and dreams) a lot, the castle is surrounded by the castles of malevolent neighbouring kingdoms. The lord tries to fortify the castle by digging deep moats around it, filling them with water and putting hungry alligators in them. But there is one thing to remember: No matter what he does to fortify the perimeter, the castle is only as strong as the resilience of the lord and the integrity of his army! Therefore, the ruler prioritises threats from the inside of his castle over those from the outside.

Please note that SEANCE Micro Business Threat Modelling Framework is not intended to be a complete cyber security solution for micro businesses, but rather to improve their cyber security posture by raising awareness of cyber threats and providing a starting point for implementing industry standard security controls.

Let's now dive into the six layers of the SEANCE framework, starting from the innermost layer and working our way outwards. You may notice that some themes, such as "Setting up firewalls", appear in more than one layer, because they are relevant to different layers.

**Layer 1: Self**

This is the innermost layer, where you consider the threats to yourself as well as the security controls you set up and manage. If your business is not wholly owned or controlled by you and you have other partners, they should also be considered in this layer.

| Key themes | Guiding questions |
|---|---|
| - Business owner(s)<br>- Administrators<br>- Organisation-wide security controls<br>- Password policy<br>- Incident Response Plan<br>- Business Continuity Plan<br>- Cyber Security Insurance | **Are you the sole owner of the business or are there other shareholders/partners who run the business with you?**<br>- How many shareholders/partners do you have?<br><br>**Do you carry out the administrative tasks yourself, or do you have a designated admin?**<br>- Do you share your admin password with anyone else?<br><br>**Do you have and enforce a password policy?**<br>- Does your password policy recommend that users create a different and unique password for each system they access? |

- Does it provide guidance to users on how to create strong passwords?

- Does it specify the technical controls for measuring and managing the quality of passwords?

- Is there a process in place to detect and change the compromised passwords?

- Are there any safeguards in place to prevent passwords from being guessed by randomly trying all possible combinations (also known as "Brute Force")?

**Do you have a policy for creating and deleting user accounts?**

- Is there a policy for creating a new user account?

- Are there any procedures in place to monitor user accounts as employees join and leave the organisation?

**Do you have a policy for managing administrator accounts?**

- Are you the only user with administrator privileges in the administration?

- If you are the only administrator in the organisation, do you share your admin account credentials with anyone else?

- Do you have a specific procedure that you follow and/or do you require an employee to go through a specific process before granting administrative privileges to an employee?

- Are there procedures to keep administrator users from using administrator accounts for everyday tasks/activities?

- Are administrator passwords protected by multi-factor authentication (MFA)?

|  | **Do you have an incident response plan in place in the event of a cyber security incident?**<br>• Are you aware of any possible legal reporting requirements and/or information sharing with the legal authorities?<br><br>**Do you have a business continuity plan to get your business back up and running as quickly as possible?**<br><br>**Do you have a cyber security insurance in place to cover or reduce the financial risks caused by cyber security events?** |
|---|---|

**Layer 2: Employees**

This layer consists of considerations related to all of your organisation's employees. This can include full-time and part-time employees, as well as volunteers, temporary or seasonal workers, etc.

| Key themes | Guiding questions |
|---|---|
| • Number of employees<br>• Remote employees<br>• User accounts<br>• User training | **How many employees do you have?**<br><br>**Do you have any remote employees who need to access to company assets remotely?**<br>• How many remote employees do you have?<br>• Do any remote employees have administrator privileges?<br>• Do remote employees change their computer/smartphone settings to stop automatic connections to public Wi-Fi?<br>• Do remote employees enable full disk encryption on the devices that they use to connect to your network? |
|  | - Do you require the employees to use secure connections when connecting remotely to your network? These measures may include:<br><br>(1) Using a router with WPA2 or WPA3 encryption when connecting from home.<br><br>(2) Using a VPN when connecting via public Wi-Fi.<br><br>**Do you regularly brief and train all employees on the procedures and processes related to the organisation's cyber security measures and policies?**<br>- Do our employees know how to recognise and report common attacks (e.g. phishing, ransomware, etc.)?<br>- Do your employees know how to perform basic cyber hygiene tasks?<br><br>**Do you have a password policy for the users?**<br>- Does it encourage/enforce users to choose unique and strong passwords for their accounts?<br>- Does it enforce that everyone uses their own account and that there are no shared accounts? |

**Layer 3: Assets**

This layer includes all the assets your organisation owns. An asset can be broadly described as something of value to your business that you want to protect. Assets include computers, servers, data stores, websites, thin clients, self-service terminals, tablets, smartphones and any other device or digital tool that is essential to the business or that are used by you and your employees.

| Key themes | Guiding questions |
|---|---|
| - Computers (desktops, laptops, thin clients, terminals, etc.)<br>- Servers<br>- Cloud services<br>- Web sites<br>- Email accounts<br>- IoT devices<br>- Anti-virus/malware protection<br>- Software/firmware updates<br>- Data security<br>- Data backups<br>- Point of Sale (POS) System | **How many computers do you use to run your business? Consider all the desktop and laptop computers that you use to store data, manage orders, and communicate with the customers.**<br><br>**Do you keep an inventory of your assets?**<br><br>- Do you keep track of the designated users and administrators of each device?<br>- Do you record the type of data that is stored on each device?<br>- Do you assess the risk of losing access to each device?<br><br>**Do you operate your own servers?**<br><br>- Do you administer the servers yourself or do you delegate this to one of your employees?<br>- Do you have procedures for who can access and configure servers and how?<br><br>**Do you use a cloud service provider to store data, run a business application, etc.?**<br><br>- Have you enabled MFA for the administrator and user accounts to access and/or manage the cloud services securely?<br>- Do you administer the cloud services yourself or do you delegate it to one of your employees?<br>- Is the administrator account shared between you and your employees, or is it kept confidential for one user?<br>- Is the administrator account used for day-to-day activities?<br>- Are you aware of the Shared Responsibility Model? |

| | **Do you use Software as a Service (SaaS) products such as Google Docs, Microsoft365, etc.?** |
|---|---|
| | **Do you use IoT devices?** |
| | • Have you changed the default passwords? Have you set strong passwords? |
| | • Do you keep the software/firmware up to date? |
| | **Have you enabled automatic updates for all of your applications, including the operating system (OS)?** |
| | • If not, do you have a process in place to apply all critical updates in a reasonable amount of time? |
| | **Are you using software that is no longer supported?** |
| | • If so, do you disconnect the devices that are running unsupported software from the Internet? |
| | **Do you review your devices regularly and disable or remove unused software?** |
| | **Are you making sure that all accounts on your devices and cloud services are only those that are used as part of your day-to-day business?** |
| | **Have you disabled Auto-Run and Auto-Play on all your devices?** |
| | **Do you lock your mobile devices with a locking mechanism like a password, fingerprint, facial recognition, etc.?** |
| | **Do you ensure that all default passwords on all devices are changed?** |
| | **Have you taken necessary measues to protect your devices (including computers, laptops, tablets and mobile devices) against malware by installing anti-malware software or only allowing a certain set of apps to be installed?** |

|  | **Do you restrict the installation of third-party (unsigned) applications outside the official application stores?** |
|--|--|
|  | **Do you keep your sensitive and critical data encrypted when it is stored (at rest) or while it is being transmitted (in transit)?** |
|  | • Do you enable full disk encryption (FDE) on your devices? |
|  | • If FDE is not enabled or available, do you encrypt sensitive files, folders, or databases individually? |
|  | • Do you use Virtual Private Network (VPN) solutions to encrypt the network traffic? |
|  | **Do you securely delete and/or destroy the data that is no longer in use?** |
|  | **Do you have a procedure in place for regularly backing up your data? If so:** |
|  | • How often do you back up your data? |
|  | • If you need to restore data from the last backup, how long does it take? |
|  | **Do you have a website? If so:** |
|  | • Do you have an SSL certificate? |
|  | • Are you using HTTPS instead of HTTP? |
|  | • Do you have a Web Application Firewall (WAF) in place? |
|  | • Do you allow users to create accounts on your website? If so: |
|  |     - Do you enforce Multi-Factor Authentication (MFA) for the users? |
|  | • Do you accept online orders on your website? |
|  | • Do you accept online payments? If so: |
|  | - Is your payment service provider PCI DSS compliant?<br><br>**Do you have procedures in place to safely dispose of and destroy devices (i.e. hard drives, flash drives, etc.) that are no longer in use?**<br><br>**Do you use a Point of Sale (POS) system? If so:**<br><br>• Is your system PCI DSS compliant?<br><br>• Do you keep the POS devices physically secure?<br><br>• Do you keep POS software and firmware up to date?<br><br>• Do you monitor POS device activity for potential suspicious activity?<br><br>• Have you changed the default passwords on your POS devices, and do you keep your passwords secure?<br><br>• Do you keep POS devices secure after hours?<br><br>• Do you have a clear policy on who can use POS devices?<br><br>• Is your POS system connected to a network to which customers and visitors can connect? |

**Layer 4: Network**

Network is the digital heartbeat of your micro business, seamlessly connecting all your devices - like computers, tablets, and smartphones - so they can talk to each other and the wider world. Think of it like a team of friends who can easily talk, share photos, or even work on projects together, regardless of where they are. This setup helps your business run more efficiently, enabling tasks like sending emails, sharing files, and accessing the internet seamlessly.

| Key themes | Guiding questions |
|---|---|
| - Firewall<br>- Routers<br>- Internet access<br>- Wi-Fi network<br>- Cloud services<br>- Shared Responsibility Model<br>- Managed Service Provider (MSP)<br>- Service Level Agreement (SLA)<br>- Network monitoring | **Is your business network protected by a firewall (or a router with firewall capability)? If so:**<br>- Have you changed all passwords from the default passwords on your firewalls and internet gateways and are all passwords difficult to guess?<br>- If you thought that passwords had been exposed (i.e. someone who knew your password had left the organisation), would you be aware of this and change the password?<br><br>**Do you have any services enabled that can be accessed externally?**<br><br>**Can your routers or hardware firewalls be configured remotely over the Internet? If so:**<br>- Have you secured the administrator account and passwords and not shared them across the organisation?<br><br>**Have you configured your routers or your hardware firewalls to block all other services being advertised to the Internet?**<br><br>**Do you keep the firmware on your router(s) updated?**<br><br>**Do you use the services of a third-party Managed Service Provider (MSP) to manage your network and network devices and to monitor the devices and networks for suspicious activity? If so:**<br>- Is there an appropriate SLA between you and the MSP?<br><br>**Do you use cloud services in your organisation? If so:** |
| | - Do you have a list of the cloud services you use?<br><br>- Have you enabled MFA for the administrator and user accounts to access and/or manage the cloud services you use?<br><br>- Have you located and understood the shared responsibility model for the cloud services you use?<br><br>**Do you allow people outside your organisation (e.g. customers, suppliers, etc.) to connect to the same Wi-Fi network that your business assets are connected to? If so:**<br>- Have you ensured that the business assets are behind a firewall? |

**Layer 5: Customers**

While customers are vital to the financial longevity of your business, they also need to be treated with the utmost care when it comes to your organisation's cyber security. As they are the most important group of people outside the organisation, you should carefully manage the relationship between customers and business assets. On the other hand, although they are external to the organisation, customers need to penetrate the external boundaries of the organisation in a structured way in order to receive services from you.

| Key themes | Guiding questions |
|---|---|
| - Privacy<br>- Wi-Fi network | **Do you collect personally identifiable information (PII) from customers for any purpose (marketing, research, etc.)?**<br><br>**Do you allow users to create and maintain accounts on your site? If so:**<br>- Do you comply with the privacy regulations depending on your jurisdiction |
| | (e.g. GDPR, DPA, etc.)?<br><br>- Do you enforce Multi-Factor Authentication (MFA) for the users?<br><br>- In the unlikely event of a customer data breach, do you have a process in place to notify and update affected customers?<br><br>**Do you allow customers to connect to the same Wi-Fi network that your business assets are connected to? If so:**<br><br>- Have you ensured that the business assets are behind a firewall?<br><br>**How do customers interact with your assets, network, employees, and yourself?** |

### Layer 6: Environment

The outermost layer refers to the external boundaries of your organisation. So far, all considerations have been about the assets and people that are part of the organisation, including customers as explained in the previous layer. This layer considers all threats and threat actors outside the organisation's perimeter.

| Key themes | Guiding questions |
|---|---|
| - Non-organisational people<br>- Firewall<br>- Physical security controls<br>- Fire safety | **Do you have a firewall enabled to filter/block incoming external traffic?**<br><br>**Approximately how many people from outside your organisation, other than customers, visit your workplace (per day/week/month/year)? Think of suppliers, government officials, etc.**<br><br>- Do you keep a record of the non-organisational visitors? |

| - Power outage<br>- Theft | - Have you noticed a sudden increase in the number of non-organisational visitors?<br>- Do you allow them to connect to your Wi-Fi network?<br><br>**Do you have physical security controls (e.g. cameras, locks, etc.) in place to secure the external perimeter of the organisation?**<br><br>**Do you have the necessary fire safety measures in place, including smoke detectors and fire extinguishers?**<br><br>**Do you have uninterruptible power supplies (UPS) in place to keep critical equipment running and prevent data loss in the event of a power outage?** |
|---|---|

# Appendix E: SEANCE Threat Modelling Tool Question Base

**Layer 1: Self**

| No | Question & sub-questions (if applicable) |
|---|---|
| 1 | Are you the sole owner of the business (e.g. do you run the business alone or with some other co-owner or a partner)?<br><br>• How many shareholders/partners do you have? |
| 2 | Do you have and enforce a password policy?<br><br>• Does your password policy recommend that users create a different and unique password for each system they access?<br>• Does it provide guidance to users on how to create strong passwords?<br>• Does it specify the technical controls for measuring and managing the quality of passwords?<br>• Does it enforce that everyone uses their own account and that there are no shared accounts?<br>• Is there a process in place to detect and change the compromised passwords?<br>• Are there any safeguards in place to prevent passwords from being guessed by randomly trying all possible combinations (also known as "Brute Force")? |
| 3 | Do you have a policy for creating and deleting user accounts?<br><br>• Is there a policy for creating a new user account?<br>• Are there any procedures in place to monitor user accounts as employees join and leave the organisation? |
| 4 | Are you the only user with administrator privileges (e.g. creating new users, setting passwords, configuring devices and network, etc.) in the business?<br><br>• How many administrator accounts you currently have?<br>• Do you have a specific procedure that you follow and/or do you require an employee to go through a specific process before granting administrative privileges to an employee? |

| | • Are there procedures to keep administrator users from using administrator accounts for everyday tasks/activities?<br>• Are administrator accounts shared across the organisation?<br>• Are administartor accounts accessed remotely?<br>• Are the administrator accounts used for day-to-day activities?<br>• Are administrator passwords protected by multi-factor authentication (MFA)? |
|---|---|
| 5 | Do you have an incident response plan in place in the event of a cyber security incident, including any possible legal reporting requirements and/or information sharing with the legal authorities? |
| 6 | Do you have a business continuity plan to get your business back up and running as quickly as possible? |
| 7 | Do you have a cyber security insurance in place to cover or reduce the financial risks caused by cyber security events? |

**Layer 2: Employees**

| No | Question & sub-questions (if applicable) |
|---|---|
| 1 | How many employees do you have (besides yourself)? |
| 2 | Do you have any remote employees who need to access to company assets remotely?<br>• How many remote employees do you have?<br>• Do any remote employees have administrator privileges?<br>• Do remote employees change their computer/smartphone settings to stop automatic connections to public Wi-Fi?<br>• Do remote employees enable full disk encryption on the devices that they use to connect to your network? |

| | • Do you require the employees to use secure connections (e.g. A router with WPA2 or WPA 3 encryption or using a VPN when connecting via public Wi-Fi) when connecting remotely to your network? |
|---|---|
| 3 | Do you regularly brief and train all employees on the procedures and processes related to the organisation's cyber security measures and policies? |
| 4 | Do your employees know how to recognise and report common attacks (e.g. phishing, ransomware, etc.)? |
| 5 | Do your employees know how to perform basic cyber hygiene tasks (e.g. choosing strong passwords, staying away from unsafe networks, not opening suspicious emails and attachments, etc.)? |
| 6 | Do your employees share accounts (i.e. they all use the same email account with same username and password instead of having separate accounts)? |

**Layer 3: Assets**

| No | Question & sub-questions (if applicable) |
|---|---|
| 1 | How many computers do you use to run your business? Consider all the desktop and laptop computers that you use to store data, manage orders, and communicate with the customers. |
| 2 | Do you keep an inventory of your assets?<br><br>• Do you keep track of the designated users and administrators of each device?<br>• Do you record the type of data that is stored on each device?<br>• Do you assess the risk of losing access to each device? |
| 3 | Do you operate your own servers? |
|   | - Are your servers behind a firewall?<br>- Do you administer the servers yourself (i.e. You DO NOT delegate this to one of your employees)?<br>- Do you have procedures for who can access and configure servers and how? |
| 4 | Do you use a cloud service provider to store data, run a business application, etc.?<br><br>- Do you administer the cloud services yourself (i.e. You DO NOT delegate this to one of your employees)?<br>- Have you enabled MFA for the administrator and user accounts to access and/or manage the cloud services securely?<br>- Are you aware of the Shared Responsibility Model? Have you fully understood your responsibilities and what to expect from the cloud service provider? |
| 5 | Do you use Software as a Service (SaaS) products such as Google Workspace, Microsoft365, Dropbox, Zendesk, etc.?<br><br>- Have you enabled multi-factor authentication (MFA) enabled for authentication to SaaS services?<br>- Do you monitor data sharing and grant appropriate permissons to each user?<br>- Do you maintain an inventory of SaaS services (i.e., who is using what service and with what permissions)? |
| 6 | Do you use IoT devices?<br><br>- Are your IoT devices behind a firewall?<br>- Do you keep the firmware of IoT devices up to date?<br>- Have you changed the default passwords of each device? Have you set strong passwords? |
| 7 | Have you enabled automatic updates for all of your applications, including the operating system (OS)?<br><br>- Do you have a process in place to apply all critical updates in a reasonable amount of time? |
| 8 | Are you using software that is no longer supported?<br><br>• Do you disconnect the devices that are running unsupported software from the Internet? |
| 9 | Do you review your devices regularly and disable or remove unused software? |
| 10 | Are you making sure that you use Identity and Access Management (IAM) accounts for day-to-day operations in cloud services, rather than admin/root accounts? |
| 11 | Have you disabled Auto-Run and Auto-Play on all your devices? |
| 12 | Do you lock your mobile devices with a locking mechanism like a password, fingerprint, facial recognition, etc.? |
| 13 | Have you changed the default passwords of each device and service? Have you set strong passwords? |
| 14 | Have you taken necessary measues to protect your devices (including computers, laptops, tablets and mobile devices) against malware by installing anti-malware software or only allowing a certain set of apps to be installed? |
| 15 | Do you restrict the installation of third-party (unsigned) applications outside the official application stores? |
| 16 | Do you keep your sensitive and critical data encrypted when it is stored (at rest) (i.e. by enabling full disk encryption (FDE) on your devices or if FDE is not enabled or available, by encrypting sensitive files, folders, or databases individually)? |
| 17 | Do you securely delete and/or destroy the data that is no longer in use? |
| 18 | Do you have a procedure in place for regularly backing up your data? |
| 19 | Do you have a website? |
| | • Is your website hosted on a remote server (i.e. on the servers of an hosting company, as opposed to being hosted on premises)?<br>• Do you have an SSL/TLS certificate (i.e. do you use HTTPS instead of HTTP)?<br>• Do you have a Web Application Firewall (WAF) in place? |
| 20 | Do you accept online payments?<br>• Is your payment service provider PCI DSS compliant? |
| 21 | Do you use a Point of Sale (POS) system?<br>• Is your system PCI DSS compliant?<br>• Do you keep the POS devices physically secure?<br>• Do you keep POS software and firmware up to date?<br>• Do you monitor POS device activity for potential suspicious activity?<br>• Have you changed the default passwords on your POS devices and do you keep your passwords secure?<br>• Do you keep POS devices secure after hours?<br>• Do you have a clear policy on who can use POS devices?<br>• Is your POS system connected to a network to which customers and visitors can connect? |
| 22 | Do you have procedures in place to safely dispose of and destroy devices (i.e. hard drives, flash drives, etc.) that are no longer in use? |

**Layer 4: Network**

| No | Question & sub-questions (if applicable) |
|---|---|
| 1 | Is your business network protected by a firewall (or a router with firewall capability)? |
| 2 | Have you changed all passwords from the default passwords on your firewalls and/or routers and are all passwords difficult to guess? |

| 3 | If you thought that passwords had been exposed (i.e. someone who knew your password had left the organisation), would you be aware of this and change the password? |
|---|---|
| 4 | Can your routers or hardware firewalls be configured remotely over the Internet?<br>• Have you secured the administrator account and passwords and not shared them across the organisation? |
| 5 | Have you configured your hardware firewalls or routers to prevent exposure of any other services on the Internet? |
| 6 | Do you keep the firmware on your router(s) updated? |
| 7 | Do you keep your sensitive and critical data encrypted while it is being transmitted (in transit) (i.e. by using Virtual Private Network (VPN) solutions, etc.)? |
| 8 | Do you use the services of a third-party Managed Service Provider (MSP) to manage your network and network devices and to monitor the devices and networks for suspicious activity?<br>• Is there an appropriate SLA between you and the MSP? |

**Layer 5: Customers**

| No | Question & sub-questions (if applicable) |
|---|---|
| 1 | Do you collect personally identifiable information (PII) from customers for any purpose (marketing, research, etc.)? |
| 2 | Do you allow users to create and maintain accounts on your website?<br>• Do you comply with the privacy regulations depending on your jurisdiction (e.g. GDPR, DPA, etc.)?<br>• Do you enforce Multi-Factor Authentication (MFA) for the users? |
|   | • In the unlikely event of a customer data breach, do you have a process in place to notify and update affected customers? |
| 3 | Do you allow customers to connect to the same Wi-Fi network that your business assets are connected to?<br><br>• Have you ensured that the business assets are behind a firewall? |

**Layer 6: Customers**

| No | Question & sub-questions (if applicable) |
|---|---|
| 1 | Do you have a firewall enabled to filter/block incoming external traffic? |
| 2 | Do you keep a record of the non-organisational visitors (e.g. government officials, suppliers, etc.) ? |
| 3 | Do you allow non-organisational visitors to connect to your Wi-Fi network? |
| 4 | Do you have physical security controls (e.g. cameras, locks, etc.) in place to secure the external perimeter of the organisation? |
| 5 | Do you have the necessary fire safety measures in place, including smoke detectors and fire extinguishers? |
| 6 | Do you train/raise awareness of your employees about the importance of physical security (e.g. such as being alert to avoid unauthorised access to sensitive locations and devices by people outside the organisation)? |
| 7 | Have you taken the necessary measures to avoid direct monitoring of sensitive business data (e.g. relocating the monitor of your admin computer so that it is not directly visible to unauthorised persons)? |

| 8 | Do you have uninterruptible power supplies (UPS) in place to keep critical equipment running and prevent data loss in the event of a power outage? |
|---|---|
| 9 | Do you use mobile devices (e.g.. laptops, tablets, removable media) in your business operations?<br><br>• Do you keep your mobile devices safe at all times? This includes keeping an eye on these devices at all times during working hours and locking them in a safe place after hours. |